\documentclass[aps,pra,preprint,amsmath,amssymb]{revtex4}
\usepackage{graphicx,epsfig,epsf}
\usepackage{dcolumn}
\usepackage{bm,psfrag,color}

\begin{document}
\def\vec#1{\mathbf{#1}}
\def\ket#1{|#1\rangle}
\def\bra#1{\langle#1|}
\def\ketbra#1{|#1\rangle\langle#1|}
\def\braket#1{\langle#1|#1\rangle}
\def\idmat{\mathbf{1}}
\def\caln{\mathcal{N}}
\def\calc{\mathcal{C}}
\def\rhon{\rho_{\mathcal{N}}}
\def\rhoc{\rho_{\mathcal{C}}}
\def\tr{\mathrm{tr}}
\def\bfu{\mathbf{u}}
\def\bfmu{\mbox{\boldmath$\mu$}}

\newcommand{\ri}{{\rm i}}
\newcommand{\re}{{\rm e}}
\newcommand{\bb}{{\bf b}}
\newcommand{\bc}{{\bf c}}
\newcommand{\bj}{{\bf j}}
\newcommand{\br}{{\bf r}}
\newcommand{\bp}{{\bf p}}
\newcommand{\bx}{{\bf x}}
\newcommand{\bz}{{\bf z}}
\newcommand{\by}{{\bf y}}
\newcommand{\bu}{{\bf u}}
\newcommand{\bv}{{\bf v}}
\newcommand{\bd}{{\bf d}}
\newcommand{\bk}{{\bf k}}
\newcommand{\bA}{{\bf A}}
\newcommand{\bB}{{\bf B}}
\newcommand{\bE}{{\bf E}}
\newcommand{\bF}{{\bf F}}
\newcommand{\bH}{{\bf H}}
\newcommand{\bR}{{\bf R}}
\newcommand{\bM}{{\bf M}}
\newcommand{\bn}{{\bf n}}
\newcommand{\bzero}{{\bf 0}}
\newcommand{\tbs}{\tilde{\bf s}}
\newcommand{\rSi}{{\rm Si}}
\newcommand{\dB}{d_{\rm Bures}}
\newcommand{\beps}{\boldsymbol{\epsilon}}
\newcommand{\bthe}{\boldsymbol{\theta}}
\newcommand{\blam}{\boldsymbol{\lambda}}
\newcommand{\bbeta}{\boldsymbol{\beta}}
\newcommand{\bxi}{\boldsymbol{\xi}}
\newcommand{\rg}{{\rm g}}
\newcommand{\xmax}{x_{\rm max}}
\newcommand{\ra}{{\rm a}}
\newcommand{\rx}{{\rm x}}
\newcommand{\rs}{{\rm s}}
\newcommand{\rP}{{\rm P}}
\newcommand{\curl}{{\rm curl}}
\newcommand{\rdiv}{{\rm div}\,}
\newcommand{\up}{\uparrow}
\newcommand{\down}{\downarrow}
\newcommand{\hc}{H_{\rm cond}}
\newcommand{\kb}{k_{\rm B}}
\newcommand{\cI}{{\cal I}}
\newcommand{\tib}{\tilde{b}}
\newcommand{\tik}{\tilde{k}}
\newcommand{\tih}{\tilde{h}}
\newcommand{\tit}{\tilde{t}}
\newcommand{\tix}{\tilde{x}}
\newcommand{\cE}{{\cal E}}
\newcommand{\cC}{{\cal C}}
\newcommand{\Ubs}{U_{\rm BS}}
\newcommand{\sinc}{{\rm sinc}}
\newcommand{\sign}{{\rm sign}}
\newcommand{\sgn}{{\rm sgn}}
\newcommand{\qq}{{\bf ???}}
\newcommand*{\etal}{\textit{et al.}}
\newcommand{\col}[1]{\textcolor{red}{#1}}

\sloppy

\title{Fourier-Correlation Imaging} 
\author{Daniel Braun$^{1}$, Younes Monjid$^2$, Bernard Roug\'e$^2$
  and Yann Kerr$^2$ } 
\affiliation{$^{1}$ Institute for Theoretial Physics, University
  T\"ubingen, 72076 T\"ubingen, Germany}
\affiliation{$^{2}$ CESBIO, 18 av. Edouard Belin, 31401 Toulouse, France}
\centerline{\today}
\begin{abstract}
We investigate to what extent  correlating the Fourier components at
slightly shifted  frequencies of the fluctuations of the electric
field measured with a one-dimensional antenna array 
on board of a satellite flying over a plane, allows one to measure
the two-dimensional brilliance temperature as function of position in
the plane.  We find that the achievable 
spatial resolution resulting from just two antennas is of the order of
$h\chi$, with $\chi=c/(\Delta r \omega_0)$, both in the 
direction of flight of the satellite and in the direction perpendicular to
it, where   $\Delta r$ is the distance between the antennas,
$\omega_0$ the central frequency, $h$ the height of the satellite over
the plane,  and $c$ the speed of light.  Two antennas  
separated by a distance of about 100m on a satellite flying with a
speed of a few km/s at a height of the order of 1000km and a central
frequency of order GHz allow therefore the imaging 
of the brilliance temperature on the surface of Earth with a resolution of
the order of one km.  For a single point source, the relative
radiometric resolution is of order $\sqrt{\chi}$, but
for a uniform temperature field in a half plane left or right  of the
satellite track it is only of order $1/\chi^{3/2}$, indicating that
two antennas do not suffice for a precise reconstruction of the temperature
field. Several ideas are discussed how the radiometric resolution
could be enhanced. In particular, having $N$ antennas all separated by
at least a distance of the order of the wave-length, allows one to
increase the signal-to-noise ratio by a factor of order $N$, but
requires to average over $N^2$ temperature profiles obtained from
as many pairs of antennas.
\end{abstract}  

\maketitle


\section{Introduction}\label{sec.intro}
Spatial aperture synthesis is a standard technique in
radio-astronomy \cite{Interferometry_Synthesis_Radio_Astronomy}.  It
allows one to achieve the fine resolution of a large 
antenna by correlating time-delayed signals
received from the different antennas in an antenna array. In
satellite-based remote sensing, spatial aperture synthesis is a
technique of choice when relatively long wave-lengths are imposed by
the applications, such as the measurement of sea surface salinity or
surface soil moisture.  When operating in the protected L-band
(1400-1427 MHz), a resolution of 10km would require already a single antenna
with a size of 32 meters.  Spatial aperture synthesis for passive
microwave-sensing was therefore
proposed to ESA \cite{Kerr2001}, and implemented for the first time
in the SMOS 
mission in 2009 that still operates today
\cite{Kerr2001b,Kerr2010}. The satellite uses a deployable Y-shaped
antenna array 
and 
provides a spatial resolution between 27-60km.\\

With the application-driven need for higher spatial resolution down to
the order of 1km, even spatial aperture synthesis leads to
forbiddingly large antenna arrays, and there is therefore an ongoing
quest for finding alternative concepts
(see e.g.~\cite{camps_two-dimensional_2001} and references therein). 
Compared to stationary  
antenna arrays on Earth used for astronomy, one may wonder whether
the motion of the satellite could be used for creating a
two-dimensional (2D) artificial antenna array out of a one-dimensional (1D)
moving array, oriented perpendicular to the motion of the
satellite. It turns out that this is not possible
when directly correlating the observed microwave fields in the
time-domain:  the useful phase-shift gained due to the motion of the
satellite is, to first order in $v_s/c$ cancelled by the Doppler
shift, where $v_s$ is the speed of the satellite and $c$ the speed of light
\cite{braun_generalization_2016}. \\ 

In this paper, we examine another idea: instead of correlating the
signals in the time domain, we consider the correlations between their
Fourier components at slightly different frequencies. This may appear
surprising at first, as, at the level of the sources, the standard
model assumption is that different frequencies
are entirely decorrelated. Nevertheless, a
hypothetical monochromatic point source is seen by different antennas
at slightly different frequencies due to 
the slightly different Doppler effect, and hence it makes sense to
correlate different frequency components from different antennas with
each other.  The useful frequency differences are tiny, down to below
one Hertz, and correspondingly long acquisition times are needed.
However, one may hope that this opens at least in principal a new way
of achieving 
a resolution of the order of a kilo-meter 
in passive microwave remote sensing in the L-band by
using the motion of the satellite for reducing a 2D antenna-array
to a 1D array. \\

We derive the principles of this 
``Fourier-correlation imaging'' (FouCoIm) technique in detail, and calculate 
the achievable spatial and radiometric resolution.  An emphasis is put
on pushing analytical calculations as far as possible, and 
testing the method at the hand of simple situations, namely a single
point source and a uniform temperature field. 
Estimation of numerical  values will be done with a standard set of
parameters: $h=700$\,km, $v_s=7$\,km/s, $\omega_0=2\pi\times
1.4$\,GHz, $T=300$\,K, $B=20$\,MHz, %
 and $\Delta r=100$\,m. This leads to
the important dimensionless parameters $\beta_s=v_s/c=2.33\cdot 10^{-5}$,
$\chi=c/(\Delta r\omega_0)=3.41\cdot 10^{-4}$, and $\tih\equiv h/
\Delta r= 7000$.

\section{Model}\label{sec.model}
We assume that the fluctuating micro-wave fields
measured at the 
position of the satellite are created by fluctuating microscopic
electrical currents at the surface of Earth
that are in 
local thermal equilibrium at absolute temperature $T(x,y)$, where $x,y$
are coordinates of a point on the surface of Earth. The entire
analysis will be in terms of classical 
electro-dynamics.  
In \cite{braun_generalization_2016} we derived the 
expression
\begin{equation} \label{ej2}
\bE(\br_1+\bv_st,t)=-\frac{\mu_0}{4\pi}\int d^3r''\frac{1}{R(t)}
\partial_{t'}\bj(\br'',t')\Big|_{t'=t-R(t)/c}\,,
\end{equation}
for the time dependent electric field arising from the current
fluctuations at the position of the satellite, 
with $R(t)=|\br_1+\bv_st-\br''|$, where $\br_1$ is the position of the antenna
at time $t=0$, $\bv_s$ the speed of the satellite in the Earth-fixed reference
frame, $\mu_0$ the magnetic permeability of vacuum, and $\bj(\br'',t)$ the
current density as function of space and time. All expressions are in the
Earth-fixed reference frame, which is more convenient for the present 
study than the satellite-fixed reference frame. It was shown in 
\cite{braun_generalization_2016} that (\ref{ej2}) is the correct far-field up to
relativistic 
corrections of the prefactor of order $\beta_s$ (due to the mixing of
electric and magnetic fields in a moving reference frame), and the
neglect of terms 
of order $\beta_s^2$ in the phase.  Eq.\eqref{ej2} does contain  in
the phase the linear Doppler 
shift and relativistic effects (including time dilation) up to order
$\beta_s$.  %
The far-field approximation is
justified for $R(t)\gg\lambda$, where $\lambda$ (of order cm in the
micro-wave regime) is the
wave-length of the radiation (see Chapt.9 in \cite{Jackson99}). \\
We substitute the Fourier-decomposition of the current density, 
\begin{equation} \label{jFT}
\bj(\br'',t)=
\frac{1}{\sqrt{2\pi}}\int_{-\infty}^\infty\,d\omega'e^{i\omega't}\tilde{\bj}(\br'',\omega')\,,    
\end{equation}
into (\ref{ej2}).  The question whether
one should differentiate $R(t)$ with respect to $t$ was answered to
the negative in \cite{braun_generalization_2016}, but it is irrelevant
if we neglect 
changes of order $\beta_s$ to the prefactor. We then find the time
dependent 
field seen by the flying antenna,
\begin{eqnarray}
\bE_{\br_1}(t)\equiv\bE(\br_1+\bv_st,t)&=&\frac{K_1}{\sqrt{2\pi}}\int
d^3r''\int 
d\omega'\frac{i\omega'}{|\br_1+\bv_st-\br''|} 
\tilde{\bj}(\br'',\omega')e^{i\omega'(t-|\br_1+\bv_st-\br''|/c)}\,,
\end{eqnarray}
with $K_1=-\mu_0/(4\pi)$. The Fourier transform of that signal is 
\begin{eqnarray}
\tilde{\bE}_{\br_1}(\omega_1)&=&\frac{1}{\sqrt{2\pi}}\int_{-\infty}^\infty
dt_1 e^{-i\omega_1 t_1}\bE_{\br_1}(t_1)\\
&=& \frac{K_1}{2\pi}\int_{-\infty}^\infty
dt_1 \int_{-\infty}^\infty d\omega'\int
d^3r''\frac{i\omega'\tilde{\bj}(\br'',\omega')}{|\br_1+\bv_st_1-\br''|} 
e^{i(\omega'-\omega_1)t_1}e^{-i\omega'|\br_1+\bv_st_1-\br''|/c}\,.\label{jtoE}
\end{eqnarray}
We assume that the current sources can be described by a Gaussian
process, where sources at different positions or different
frequencies, or with different polarizations are uncorrelated,   
\begin{equation} \label{curcor0}
\langle
\tilde{j}_i(\br_1'',\omega_1)\tilde{j}_j^{*}(\br_2'',\omega_2)\rangle
=\delta_{ij}\frac{l_c^3}{\tau_c}\delta(\br_1''-\br_2'')\delta(\omega_1-\omega_2)\langle|\tilde{j}_i(\br_2'',\omega_2)|^2\rangle\,,  
\end{equation}
where we have introduced for dimensional grounds a correlation length $l_c$
and a correlation time $\tau_c$, and the polarizations are indexed by $i,j$,
taking values $x,y,z$. In principle the average
$\langle\ldots\rangle$ is over an ensemble of realizations of the stochastic
process, but we may
assume ergodicity of the fluctuations, such that they can also be obtained
from a sufficiently long temporal average.  In practice this means that one
should average over positions considered as equivalent in terms of the
ensemble, i.e.~the time the satellite  takes to fly over a desired pixel
size.  For a satellite flying at a speed of order km/s and a pixel size of
order km, this means a maximal averaging time of the order of a
second. This does not preclude calculating Fourier
transforms with finer spectral resolution from data acquired over much
longer times.  \\ 

We will
make the assumption that only the current intensities at the surface of
Earth contribute.  In reality the emission seen by the satellite
arises from a thin surface layer on Earth that has a finite thickness
$d$ of
the order of a few centimeters \cite{Kerr2010,Kerr2012}, depending on
the soil and its humidity, and the
satellite also sees 
the cosmic microwave background. We approximate the surface layer as a
 single plane located at $z''=0$,
i.e.~$\langle|\tilde{j}_i(\br_2'',\omega_2)|^2\rangle 
=d\langle|\tilde{j}_i(x'',y'',\omega_2)|^2\rangle \delta(z'')$ and
neglect the cosmic 
microwave background as its temperature is two orders of magnitude
lower than that of Earth, as well as other astronomical objects.\\ 
      
The current intensities are related to an effective temperature
$T(x,y)$ by 
\begin{equation}
  \label{eq:jtoT}
\langle|\tilde{j_i}(x,y,\omega)|^2\rangle=K_2 T(x,y)\,, 
\end{equation}
where $K_2$ is a constant (see
eq.(\ref{eq:jtoT2})). Eq.(\ref{eq:jtoT}) is valid   
for 
$\hbar\omega\ll k_BT$ and hence very well adapted to 
micro-wave emission at room temperature. 

Eq.\eqref{curcor0} together with \eqref{eq:jtoT} is a standard model
of classical white noise 
currents, and appears in many places in the literature, see
e.g.~eq.(4.16) in \cite{sharkov_passive_2003}. 
The equation is an instance of the 
fluctuation-dissipation theorem that can be found in standard
text-books on statistical physics (see e.g.~Part 1, Chap. XII and Part 2,
Chap.VIII.~in\cite{LLStatMech80}). In the context of 
thermal radiation it goes back at least to the original Russian version 
of \cite{rytov_theory_1959} (from 1953); see also \cite{Rytov89}.  The
model has also been used to study coherence effects in the thermal radiation of
near-fields (see eq.(3) in \cite{carminati_near-field_1999}).  
For
completeness, we present the derivation of \eqref{eq:jtoT}  in the
appendix,  based on  
Planck's law for the energy density of an e.m.~field in thermal
equilibrium.\\

Compared to a black body, the emissivity of a real body is
modified by a mode-dependent emissivity factor
$B_i(x,y;\omega,\hat{k})$, where $\hat{k}$ is the direction of
emission (from the patch on ground to the satellite), and a factor
$\cos\theta(x,y,h)$ of geometrical origin that takes into account the 
variation of the radiation with respect to the
surface normal (i.e.~the projection of the area of a patch of the
surface onto the plane perpendicular to the propagation
direction).  The temperature $T(x,y)$ is then really an effective
temperature, $T_\text{eff}(x,y)=
T_B(x,y)\cos\theta(x,y,h)$, where   
the brightness temperature $T_B(x,y)$
is defined as
the absolute temperature a black-body would need to have in order to produce
the same intensity of radiation at the frequency and in the direction
considered (see Appendix \ref{sec:thrad}). 
For simplifying notations,  in the following we keep writing $T(x,y)$
for short instead of $T_\text{eff}(x,y)$, but keep in mind its physical
meaning, which, after all, is crucial for
data-analysis and fitting vegetation and 
surface models to observational data \cite{Kerr2012}.   
We thus arrive at the current correlator
\begin{equation} \label{curcor}
\langle
\tilde{j}_i(\br_1'',\omega_1)\tilde{j}_j^*(\br_2'',\omega_2)\rangle
=\delta_{ij}K_3\,\delta(\br_1''-\br_2'')\delta(\omega_1-\omega_2)
T(x'',y'')\delta(z'')\,,
\end{equation}
which can be considered the statistical model underlying the imaging
concept, and $K_3=l_c^3 K_2d/\tau_c$.

\section{Correlation of Fourier components}
For each antenna, the electric field component
$E_{i,\br_1}(t)$ is transduced into a voltage
$U_{i,\br_1}(t)$. We denote the frequency 
response of the antennas and 
eventual subsequent filters by the complex function $A(\omega)$, the
Fourier transform of the time-dependent response function of antenna and
filter.  In the 
frequency domain we have simply
$\tilde{U}_{i,\br_1}(\omega_1)=A(\omega_1)\tilde{E}_{i,\br_1}(\omega_1)$. 
With (\ref{curcor}) we obtain the correlation function between the voltages at 
two different frequency components $\omega_1$, $\omega_2$ measured at the
positions of the antennas with original positions $\br_1$ and $\br_2$,
\begin{eqnarray}
C_{ij}^F(\br_1,\br_2,\omega_1,\omega_2)&\equiv& \langle
\tilde{U}_{i,\br_1}(\omega_1)\tilde{U}_{j,\br_2}^*(\omega_2)\rangle =C_{ij}(\br_1,\br_2,\omega_1,\omega_2)A(\omega_1)A^*(\omega_2)\label{CijF}\\
C_{ij}(\br_1,\br_2,\omega_1,\omega_2)&=&\langle
\tilde{E}_{i,\br_1}(\omega_1)\tilde{E}_{j,\br_2}^*(\omega_2)\rangle\\
&=&K_5\delta_{ij} \int_{-\infty}^\infty
dt_1\int_{-\infty}^\infty
dt_2 \int_{-\infty}^\infty d\omega'\int
dx''dy''\frac{\omega'^2T(x'',y'')}{|\br_1+\bv_st_1-\br''||\br_2+\bv_st_2-\br''|}\nonumber\\  
&&\times
e^{i\omega'(t_1-t_2)}e^{-i(\omega_1t_1-\omega_2t_2)}e^{-i\frac{\omega'}{c}(|\br_1+\bv_st_1-\br''|-|\br_2+\bv_st_2-\br''|)}\,,  \label{C}
\end{eqnarray}
where now $\br''=(x'',y'',0)$, and $K_5=K_3K_1^2/(4\pi^2)$. 
 The correlation function
$C_{ij}^F(\br_1,\br_2,\omega_1,\omega_2)$ is the filtered version of the
original unfiltered correlations $C_{ij}(\br_1,\br_2,\omega_1,\omega_2)$. We
see from (\ref{CijF}) that the latter can be obtained from the former simply
by dividing through the product of the known filter functions, as long as
the latter are non-zero.  Of course, outside the frequency response of the
antennas and filters, the measured correlations
$C_{ij}^F(\br_1,\br_2,\omega_1,\omega_2)$ vanish due to the vanishing of
$A(\omega)$ and do not carry any information anymore. This will ultimately limit 
the frequency range over which information on the brightness
temperature can be extracted, or, equivalently, leads to a finite
geometrical resolution even if a $C_{ij}(\br_1,\br_2,\omega_1,\omega_2)$
known for all frequencies would lead to perfect resolution. However, this
appears only when inverting the measured signals and will be
discussed in section \ref{sec.inv}. For the moment we
assume that we have access to the unfiltered 
$C_{ij}(\br_1,\br_2,\omega_1,\omega_2)$ through (\ref{CijF}) for all
frequencies that we need, and base the general development of the theory
on $C_{ij}(\br_1,\br_2,\omega_1,\omega_2)$. \\

We change integration variables from $t_1,t_2$ to ``center-of-mass''
and relative times, $t=(t_1+t_2)/2$ and 
$\tau=(t_2-t_1)$, and introduce as well a new integration variable for 
the spatial 
integration, $\br'\equiv \br''-\bv_st$. This implies $\br_1+\bv_st_1-\br''=
\br_1-\bv_s\tau/2-\br'$ and $\br_2+\bv_st_2-\br''=
\br_2+\bv_s\tau/2-\br'$. The 
Jacobian of both transformations is equal to  
1. Furthermore, from now on we take
the satellite to move in $x$ direction, $\bv_s=v_s\hat{e}_x$, where
$\hat{e}_x$ is the unit vector in $x$ direction. This leads to
$T(x'',y'')=T(x'+v_st,y')$.

The total phase $\Phi$ appearing as arguments of 
the exponential functions under the
integrals in (\ref{C}) is
\begin{equation} \label{phi}
i\Phi=i\left[
\tau\left(-\omega'+\frac{\omega_2+\omega_1}{2}\right)+t(\omega_2-\omega_1)-\frac{\omega'}{c}\left(|\br_1-\bv_s\tau/2-\br'|-|\br_2+\bv_s\tau/2-\br'|\right) 
\right]\,.
\end{equation}
We see that $t$ only appears as prefactor of $(\omega_2-\omega_1)$ in the
phase (\ref{phi}), and as argument $\bv_s t$ in $T(x'+v_x t,y')$.  The
integral over $t$ therefore boils down to a 1D Fourier transform of the
intensity of the current fluctuations in the direction of the speed 
of the satellite, with conjugate variable proportional to the difference
$\omega_2-\omega_1$ of
the frequencies of the Fourier components that we correlate.
This can be made more explicit by introducing a position variable $x=v_st$
along the path of the satellite. For the conjugate variable
we define $\kappa_x=(\omega_2-\omega_1)/v_s$.  We write $\kappa_x$ and not
$k_x$ in order to distinguish this ``wavevector'' from the usual one
obtained from a single frequency and dividing by $c$.  We also
introduce the ``center of mass frequency'' 
$\omega_c\equiv(\omega_1+\omega_2)/2$.  It will be called ``center
frequency'' in the following for short, but should not be confused with the 
central frequency $\omega_0$ that is the fixed frequency in the middle of
the band in which the satellite operates (e.g.~$2\pi\times $1.4\,GHz for
SMOS). With all  this, we see that    
\begin{eqnarray} \label{Ii}
\int T(x'+v_st,y')e^{i(\omega_2-\omega_1)t}dt&=&\frac{1}{v_s}\int
T_{\br'}(x)e^{i\kappa_x
  x}\,dx=\frac{\sqrt{2\pi}}{v_s}\tilde{T}_{\br'}(\kappa_x)\\
&\equiv&\frac{\sqrt{2\pi}}{v_s}\tilde{T}_{x',y'}(\kappa_x)\,. 
\end{eqnarray}
We have defined $T(x'+v_st,y')\equiv T_{\br'}(x)$, where
$v_st=x$ is understood, and the 
spatial Fourier transform $\tilde{T}_{\br'}(\kappa_x)$ of the
temperature field $T(x,y)$  in $x$-direction.  This notation makes clear that
in these coordinates the temperature depends both on $\br'$ and $t$,
even though the motion of the 
satellite combines the two arguments in a single one, $\br'+\bv_st$.  We can
think of  $\tilde{T}_{\br'}(\kappa_x)$ as the Fourier image of
$T(\br'+x\hat{e}_x)$ with respect to the $x$ coordinate, calculated with a
starting point $\br'$. I.e.~for all $\br'$, we have a 1D spatial Fourier
transform of the intensity of the current fluctuations where the
Fourier integral is defined with origin in $\br'$.  The Fourier
images obtained by translation of $\br'$ in $x$-direction are not
independent. Rather we have 
\begin{eqnarray}
\tilde{T}_{x',y'}(\kappa_x)&=&\frac{1}{\sqrt{2\pi}}\int dx T_{x',y'}(x)e^{i\kappa_x
  x}=\frac{1}{\sqrt{2\pi}}\int dx T_{0,y'}(x+x')e^{i\kappa_x 
  x} \\
&=& \frac{1}{\sqrt{2\pi}}\int dx'' T_{0,y'}(x'')e^{i\kappa_x
  x''}e^{-i\kappa_x x'}=e^{-i\kappa_x x'}\tilde{T}_{0,y'}(\kappa_x)\,. \label{Ix0}
\end{eqnarray}
We are thus led to
\begin{eqnarray}
C_{ij}(\br_1,\br_2,\omega_1,\omega_2)&=&K_5\delta_{ij}\frac{\sqrt{2\pi}}{v_s} \int_{-\infty}^\infty
d\tau\int_{-\infty}^\infty
d\omega' \omega^{'2}\int
dx'\,dy'\frac{\tilde{T}_{0,y'}(\kappa_x)e^{-i\kappa_x x'}}{|\br_1-\bv_s\tau/2-\br'||\br_2+\bv_s\tau/2-\br'|}\label{C2}\\  
&&\times
\exp\left[i\left(\tau(-\omega'+\frac{\omega_2+\omega_1}{2})-\frac{\omega'}{c}(|\br_1-\bv_s\tau/2-\br'|-|\br_2+\bv_s\tau/2-\br'|)\right)\right]\,.\nonumber 
\end{eqnarray}
We neglect the slow dependence of $\omega'^2$  
compared to the rapid oscillations of the phase factors in (\ref{C2}) and pull it
out of the integral as a prefactor $\omega_0^2$.
We can then perform the
integral over $\omega'$, and find 
\begin{equation} \label{inin}
\int_{-\infty}^\infty\exp\left[\ldots\right]d\omega'=2\pi\delta(\tau+\frac{1}{c}(|\br_1-\bv_s\tau/2-\br'|-|\br_2+\bv_s\tau/2-\br'|))\,e^{i\tau\frac{\omega_1+\omega_2}{2}}\,.   
\end{equation}
We introduce center-of-mass and relative coordinates for $\br_1$ and
$\br_2$, $\bR=(\br_1+\br_2)/2$ and $\Delta \br=\br_2-\br_1$. We
further restrict $v_s\tau$ to values much smaller than $|\bR-\br'\pm \Delta
r|$. This implies a limitation of the integration range for $\tau$ when
calculating the Fourier components, but it is a mild one.  Since
$|\bR-\br'\pm \Delta 
r|\ge h$, it is enough to have $\tau\le h/v_s$, which is typically of order
$100$s, and therefore gives time to resolve Fourier components down to a
hundredth of a Hertz.  We can then approximate
to first order in $v_s$,  
\begin{equation} \label{1st}
|\br_1-\bv_s\tau/2-\br'|-|\br_2+\bv_s\tau/2-\br'|\simeq
 -\hat{e}_{\bR-\br'}\cdot (\Delta\br +\bv_s \tau)\,.
\end{equation}
Neglecting terms of order $|\Delta \br +\bv_s\tau|/|\bR-\br'|$ and
of order $\beta=v_s/c$ in the prefactor %
of the exponential, as well as a second order term of order
$\beta\omega_c\Delta r/c $ in the 
phase, the integral over the Dirac $\delta$-function gives 
\begin{eqnarray}
C_{ij}(\br_1,\br_2,\omega_1,\omega_2)&=&K_6\delta_{ij} \int_{-\infty}^\infty
dx'\int_{-\infty}^\infty
dy'\frac{\tilde{T}_{0,y'}(\kappa_x)e^{-i\kappa_x x'}}{|\bR-\br'|^2}\nonumber\\   
&&\times
\exp\left[i\frac{\Delta \br\cdot\hat{e}_{\bR-\br'}}{c}\omega_c \right]\,,
\label{C3} 
\end{eqnarray}
and
$K_6=(2\pi)^{3/2}\omega_0^2 K_5/v_s$.  
The unit vector $\hat{e}_{\bR-\br'}$ is obtained by taking the original
center of 
mass position of the antennas at $\bR=(x_0,0,h)$, and
$\br'=(x',y',0)$. Eq.(\ref{C3}) is one of the central results of this
paper. It shows that the two-frequency
correlation function of the fields at different antenna positions is related
linearly via a 2D integral-transformation to the brightness temperature
field in the
source plane, or more precisely to the Fourier transform of that
field in $x$-direction.  With $T(x,y)$ defined on a 2D
grid, the  
reconstruction of $T(x,y)$  from the measured correlation function
thus becomes a 
matrix inversion problem that has to be performed numerically in
general.    A crucial question is the conditioning of the
inversion problem.  It will be studied in more detail in a
subsequent paper dedicated to a numerical approach \cite{NumPaper}.

Here we give a simplified analytical treatment that allows us to
obtain estimates of the spatial 
and radiometric resolutions, and thus provide
evidence that the inversion problem is sufficiently well
conditioned for the reconstruction of $T(x,y)$ from the measured
$C_{i,j}(\br_1,\br_2,\omega_1,\omega_2)$.   For this, 
we study the situation where the vector $\Delta\br$ from
antenna 1 to antenna 2  is orientated in $y'$ direction, $\br_2=\br_1+\Delta
r\hat{e}_y$, 
in which case $\Delta\br\cdot\hat{e}_{\bR-\br'}=-\Delta r
y'/\sqrt{(x'-x_0)^2+y'^2+h^2 }$, and $\Delta r=|\Delta\br|$ denotes the spatial
separation of the two antennas.  \\

We switch to a dimensionless
representation by taking as length scale the
distance $\Delta r$ between the two antennas.  We will express all other
lengths in this unit, and introduce the dimensionless coordinates
$\xi,\eta$ by $x'=\xi \Delta r$, $y'=\eta \Delta r$, and $\tih\equiv
h/\Delta r$. The
dimensionless height $\tih$ is for the standard parameters 
$\tih=7\cdot 10^3$. 
Eq.(\ref{C3}) then reads  
\begin{equation} \label{Cfin}
C_{ij}(\br_1,\br_1+\Delta r\hat{e}_y,\omega_1,\omega_2)=K_6\delta_{ij}e^{-i\kappa_x
  x_0}\int_{-\infty}^\infty 
\frac{d\eta}{\sqrt{\eta^2+\tih^2}} K(\kappa_x\Delta
r\sqrt{\eta^2+\tih^2},\frac{\Delta
  r\omega_c}{c}\frac{\eta}{\sqrt{\eta^2+\tih^2}})\tilde{T}(\kappa_x,\eta)
\,, 
\end{equation}
where $\tilde{T}(\kappa_x,\eta)\equiv \tilde{T}_{0,\eta\Delta
  r}(\kappa_x)$. 
The 1D integral kernel  
\begin{equation} \label{kern2}
K(\alpha,\beta)=\int_{-\infty}^\infty d\xi \frac{e^{-i (\alpha
    \xi+\frac{\beta}{\sqrt{\xi^2+1}})}}{\xi^2+1}\,,
\end{equation}
which is itself defined through an integral over $\xi$. 
For fixed $h$, $\Delta r$, $\omega_c$, and $\kappa_x$, the integral
kernel $K(\alpha,\beta)$ is a function of $\eta$ that relates the
1D Fourier transform $\tilde{T}(\kappa_x,\eta)$ to the
observed correlation function by integration over $\eta$.
Suppose that the integration over $\eta$ can be inverted by finding the inverse
integral kernel. Integrating the inverse kernel  over with the
correlation function 
measured as function of the center frequency $\omega_c$, we then obtain
$\tilde{T}_{0,\Delta r
  \eta}(\kappa_x)$ for all $\eta$ and 
the  chosen $\kappa_x$. If this can be done for all relevant $\kappa_x$, we
obtain for each point on the $y$ axis the Fourier transform in $x$
direction of the intensity of the brightness temperature.  %
Taking the
inverse 
Fourier transform in $x$-direction, we obtain the full $x$-
and $y$-dependent brightness temperatures. To proceed, we first study
the integral kernel in the relevant parameter regimes. \\

\section{Properties of the integral kernel}\label{propK}
The arguments
$\alpha,\beta$ of $K$ are given by eq.(\ref{Cfin}) as 
 \begin{eqnarray}
 \alpha&=&\kappa_x\Delta r\sqrt{\eta^2+\tih^2}\\
 \beta&=&\frac{\Delta r
   \omega_c}{c}\frac{\eta}{\sqrt{\eta^2+\tih^2}}\,.\label{beta} 
 \end{eqnarray}
 By their definition, we only need 
$\alpha,\beta\in\mathbb{R}$.  For   
$\alpha$ we can consider that in the end the maximum $\kappa_x$ should be
of the order of the inverse resolution $\Delta x_{\rm min}$ required in $x$
direction. Taking  
$\Delta x_{\rm min}$ of the order of one km, and using the standard
parameters, we get $|\alpha|_{\rm max}\ge|\kappa_xh|\simeq
700$. With $\eta$ 
varying from $-\infty\ldots\infty$ (in reality, the
extension of Earth limits the integration range to a maximum value
of the order $10^7-10^8$), $\beta$ reaches
its maximal value $\Delta r\omega_c/c$ for
$\eta\to\infty$. For standard parameters, $\omega_c=2\pi\times
$1.4\,GHz, $|\beta|\lesssim 30$.  Both $\alpha$ and $\beta$ can be
positive or negative, such that there is also a regime, where
$|\beta|\gg|\alpha|$, and we will find that this is the most important
one. Note that from (\ref{kern2}) we immediately obtain
 the relations
\begin{eqnarray}
K(\alpha,\beta)&=&K(-\alpha,\beta)=K(\alpha,-\beta)^*\,.\label{sym}
\end{eqnarray}
We therefore restrict the following discussion to $\alpha,\beta\ge 0$.\\
 
Unfortunately, the integral over $\xi$ in (\ref{kern2})
cannot be done 
analytically. However, we can find approximations for different
cases. Consider first $\beta=0$. Using the methods of residues, one easily
finds 
\begin{equation} \label{b0}
K(\alpha,0)=\pi e^{-\alpha}\,.
\end{equation}
More generally, one can obtain a useful expansion for small $\beta$ by
expanding $\exp(-i \beta/\sqrt{\xi^2+1})$ into a power series, and then
integrating term by term. We find 
\begin{eqnarray}
K(\alpha,\beta)&=&\sum_{n=0}^\infty\frac{1}{n!}\int_{-\infty}^\infty (-i
\beta)^n\frac{e^{-i\alpha\xi}}{(\xi^2+1)^{1+\frac{n}{2}}}\,d\xi\\
&=&\sqrt{2\pi\alpha}\sum_{n=0}^\infty
\frac{\left(-i\beta\sqrt{\alpha/2}\right)^n}{n!\Gamma(1+n/2)}K_{(n+1)/2}(\alpha)\,,  \label{Iasmall}
\end{eqnarray}
where $K_n(x)$ is the modified Bessel function of the second kind of order
$n$.  The zeroth order result 
(\ref{b0}) is recovered by observing that
$K_{1/2}(x)=\sqrt{\frac{\pi}{2x}}e^{-x}$.  
For small $\beta$ the
series converges rapidly, and one can even improve the agreement with the 
numerically calculated kernel by re-exponentiating the first few terms.  For
example, up to second order we have a polynomial $p_0+p_1\beta+p_2\beta^2$
which we wish to write as $p_0\exp(a_1\beta+a_2\beta^2)$. Expanding the
exponential  
in powers of $\beta$ and comparing powers up to order $\beta^2$,
one finds $a_1=p_1/p_0$ and $a_2=p_2/p_0-(p_1/p_0)^2/2$. When plotted
together with the exact result, the thus obtained
approximation agrees with $K(\alpha,\beta)$ for $\alpha=2$ visibly well up
to $\beta\simeq 4$, i.e.~well beyond the regime $\beta\ll 1$. For the fourth
order re-exponentiated form the agreement extends up to about $\beta\simeq
5$.  However, the exponential decay (\ref{b0}) already of the zeroth order
term with 
$\alpha$ indicates that for the values of $\alpha\simeq 10^2$ to $10^3$ the
contribution to the $\eta$ integral for values such that $\beta$ is of order
of or smaller than one, can be entirely neglected. 

In the opposite regime of large $\beta$, an approximation based on a
stationary phase approximation can be found.  More precisely, one needs
$\beta\gg \alpha$. In that case one can treat $e^{-i\alpha \xi}/(\xi^2+1)$
as a slowly varying factor compared to the rapidly oscillating
$e^{-i\beta/\sqrt{\xi^2+1}}$.   The point of stationary phase of the latter
term is found at $\xi=0$ (where the function has a maximum). The second
derivative of the phase at $\xi=0$ equals 1. With this we get 
\begin{equation} \label{Iasy}
K(\alpha,\beta)\simeq \sqrt{\frac{2\pi}{\beta}}e^{i\pi/4}e^{-i\beta}\,,
\end{equation}
valid for $\beta/\alpha\gg 1$.  
Interestingly, the integral kernel becomes independent of $\alpha$ in this
regime, which is of course a consequence of the fact that the stationary
phase point is at $\xi=0$, thus eliminating the factor $\alpha$ in the
phase of the prefactor.  We furthermore see that in this regime there is no
exponential suppression of the kernel. \\
For $\alpha=0$, the kernel can be evaluated exactly, 
\begin{equation}
  \label{eq:K0}
  K(0,\beta)=\int_{-\infty}^\infty
  d\xi\frac{e^{-i\frac{\beta}{\sqrt{\xi^2+1}}}}{\xi^2+1}=\pi (J_0(\beta)-iH_0(\beta))\,,
\end{equation}
where $J_0$ is the zeroth Bessel function, and $H_0$ the zeroth Struve
function.  Their asymptotic behavior gives back \eqref{Iasy}.\\

\begin{figure}
 \includegraphics[width=7cm]{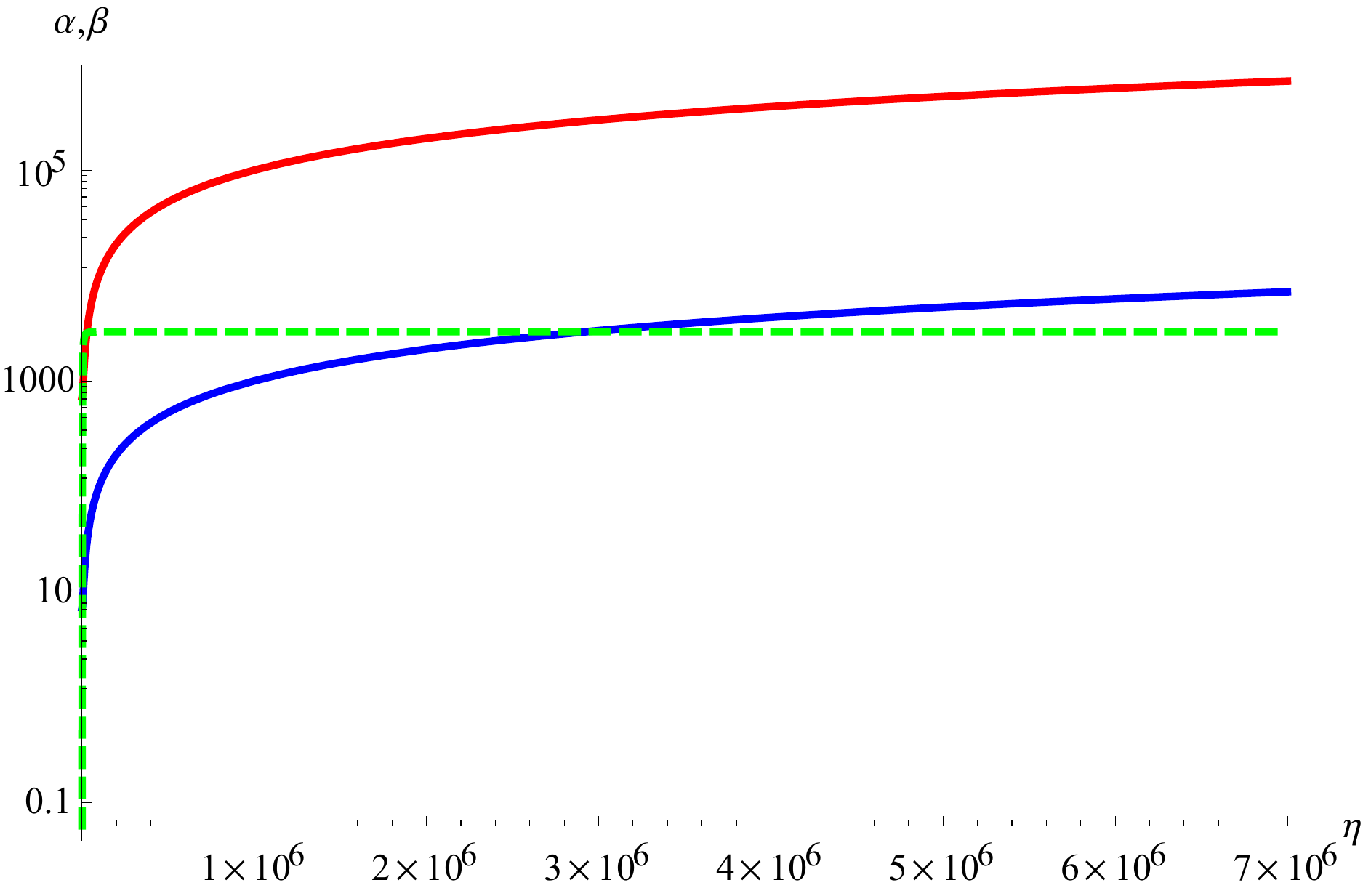}
 \includegraphics[width=7cm]{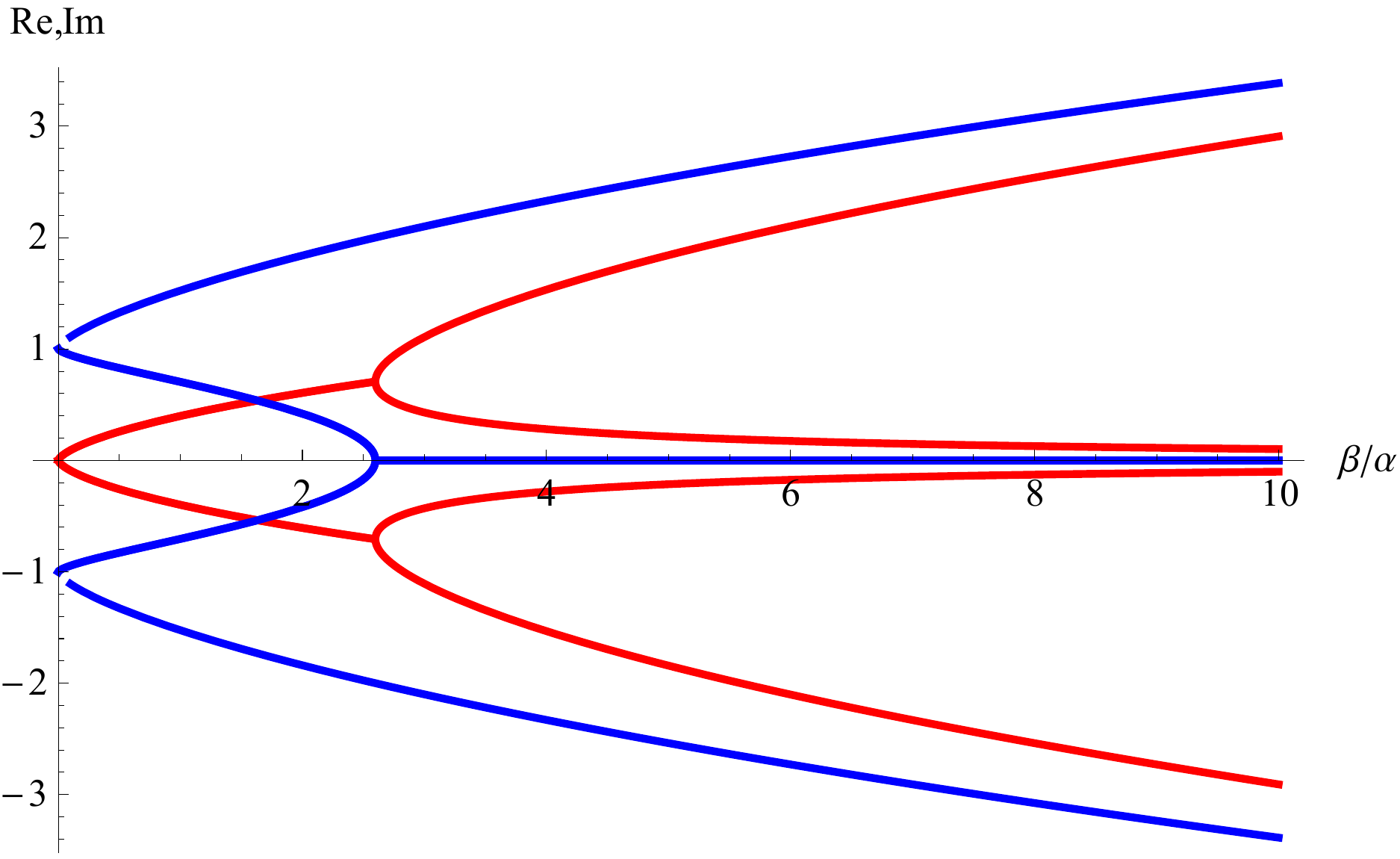}
\caption{(left) $\alpha$ and $\beta$ as function of $\eta$. Standard
  parameters are used (see sec.\ref{sec.intro}), and two different
  values for $\kappa_x$: 
  $\kappa_x=10^{-5}/$m (blue curve for $\alpha$) and
  $\kappa_x=10^{-3}/$m (red curve for $\alpha$); $\beta$ (green dashed
  curve) is independent of $\kappa_x$. (right) Real (red) and
  imaginary parts (blue) of the roots of the stationary phase equation in the
regime $\alpha\sim\beta$ as function of $\beta/\alpha$.}
\label{fig:abofeta}
\end{figure}

In Fig.\ref{fig:abofeta} we plot $\alpha,\beta$ as function of $\eta$. We
see that a regime $\alpha<\beta$ exists for 
$\kappa<\kappa_{\rm Max}$, which defines $\kappa_{\rm Max}$ (see after
eq.(\ref{eta12}) for its precise value).  For $\kappa\ll\kappa_{\rm
  Max}$, $\alpha\ll\beta$.  The regime 
$\alpha\sim\beta\ll 1$ is also possible, but it is restricted to a
tiny $\eta$ interval, such that its contribution to the integral over
$\eta$ is negligible. For $\alpha\sim\beta\gg 1$, the
stationary phase points of $\xi+(\beta/\alpha)/\sqrt{\xi^2+1}$ become
relevant. Fig.\ref{fig:abofeta} shows the Im- and
Re-parts of the six roots of 
the corresponding stationary phase equation.  We see that only for
$(\beta/\alpha)\gtrsim 2.5$ 
real stationary phase points exist. Since
on the other hand  $\alpha\sim\beta\gg 1$ occurs for sufficiently
large $\kappa$ for almost all $\eta$ only for $\beta<\alpha$ 
(see
Fig.\ref{fig:abofeta}), 
the kernel is exponentially small in this regime $\alpha\sim\beta\gg
1$. Altogether, the only relevant regime is thus $\beta\gg\alpha\gg 1$. \\
While the asymptotic form of the integral kernel suggests the use of the
orthogonality relations of Bessel functions, inverting (\ref{C3}) is
nevertheless non-trivial due to the more complicated dependence of $\alpha$
and $\beta$ on $\eta$.   However, the above
asymptotic form allows one to obtain an 
approximate analytical inversion of the kernel that allows for an
estimation of the resulting resolution, as we will show  now. 

\section{Estimation of geometrical resolution}
\subsection{Approximate analytical inversion of the integral kernel}\label{sec.inv}
\begin{figure}
(a)\includegraphics[width=6cm]{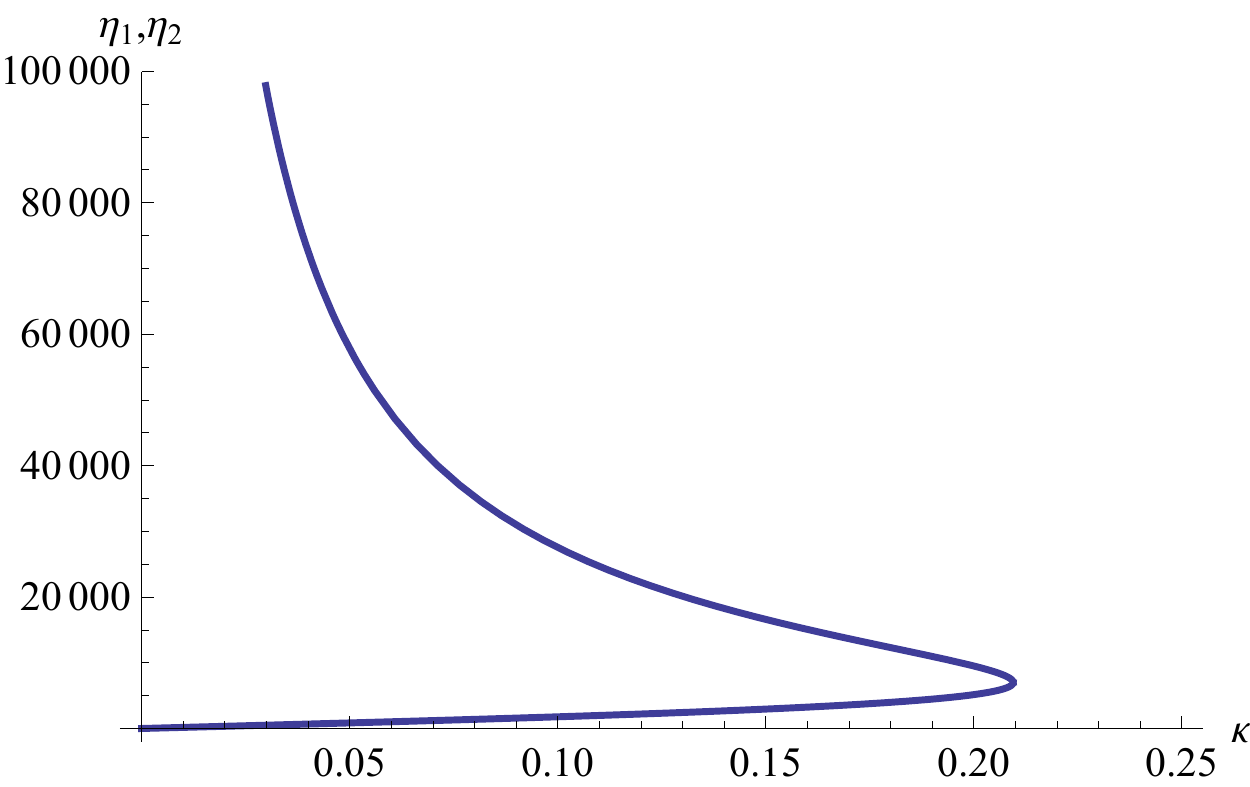}
(b)\includegraphics[width=6cm]{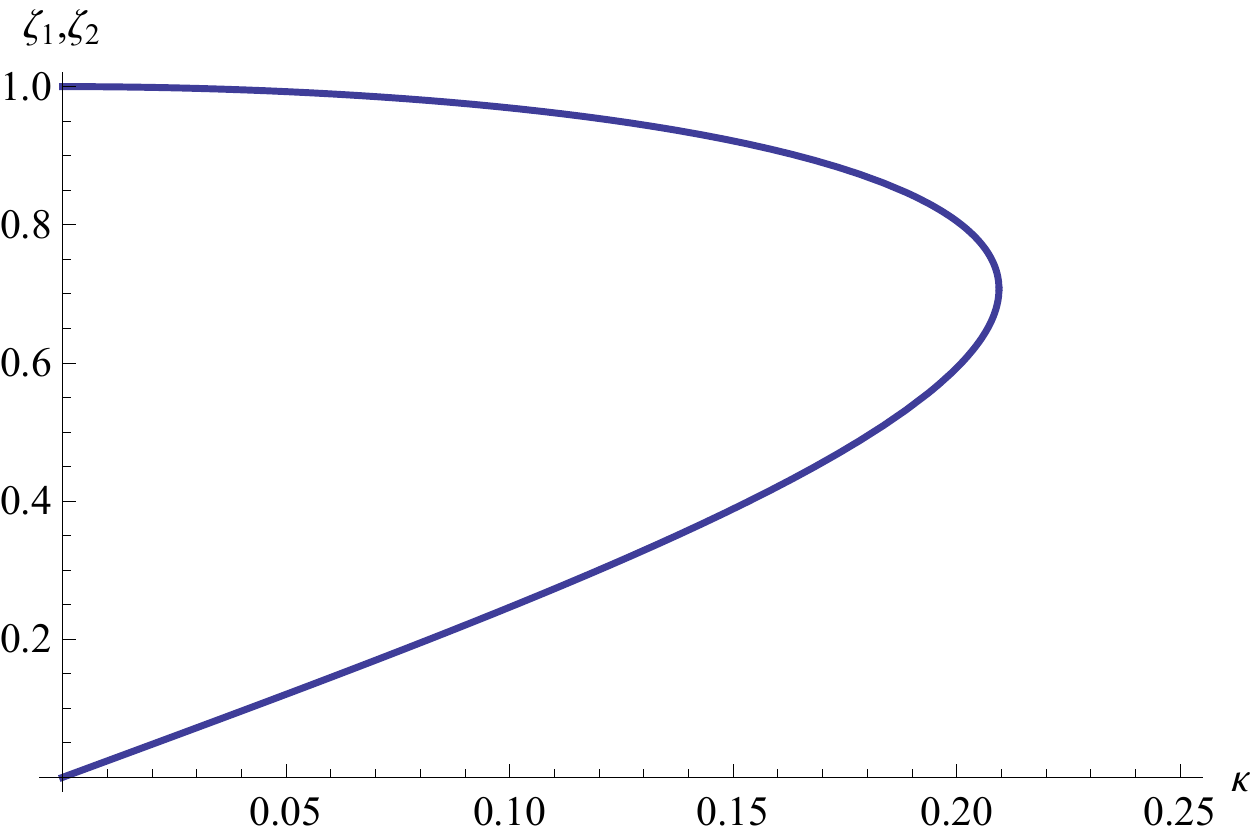}
\caption{Effectively contributing integration region as function of
  $\kappa=\kappa_x \Delta r$: (a) in terms of 
  $\eta$ and (b) in terms of $\zeta$.  Only the area in the $xy$-plane
  between the two curves, $\eta_1\le\eta\le\eta_2$ and correspondingly
  for $\zeta$, contributes effectively to the
  correlation function for a given value of $\kappa$. The two curves
  join at $\kappa_{\rm Max}\simeq 0.21$ (numerical value for standard
  parameters, see sec.\ref{sec.intro}).  Only 
  the region with 
  $\kappa,\eta\ge 0$ is shown; three more regions contribute in the
  other three quadrants, and the boundaries are obtained by reflecting
  the graph at the $\eta$-axes and $\kappa$-axes. The integration
  region translates directly into the area 
  ``seen'' by the satellite in $y$-direction for a given wave vector
  $\kappa$ in $x$-direction. For $\kappa\to 0$, the integration region
  is in reality cut-off by the size of Earth, and the smallest value of
  $\kappa$ is determined by the desired size of the snapshot or the
  maximum %
 time $\tau<h/v_s$. }
\label{fig:eta12}
\end{figure}

At first sight, the requirement
$\beta\gg \alpha$ appears unnatural given that $\alpha$ can already of the
order $10^2$ to $10^3$. And indeed, this leads to a first rather stringent
condition which must be met in order for the correlation function $C$ to be
non-zero. In terms of the original parameters, $\beta/\alpha=\omega_c
\eta/(c \kappa_x (\eta^2+\tih^2))$. 
For this to be much larger 
than one, one needs  
\begin{equation} \label{thresh}
\frac{\eta}{\eta^2+\tih^2}\gg \frac{c
  \kappa_x}{\omega_c}=\frac{c}{v_s}\frac{\Delta \omega}{\omega_c}\,,
\end{equation}
or $\Delta\omega/\omega_c\ll (v_s/(2c \tih))$, where we have used already the
maximum value $1/(2\tih)$ of the function of $\eta$ on the left hand
side (lhs) in
(\ref{thresh}). For the standard parameters, we find
$\Delta\omega/\omega_c\ll 1.66\cdot 10^{-9}$. 
When operating at $\omega_c$ in the GHz regime, this means that
the correlation function 
essentially vanishes for $\Delta\omega$ larger than a few Hertz  and thus
bears no more information for the measurement of the position
dependent brightness temperature.\\
Another way of seeing this is to observe that (\ref{thresh}) limits the
integration range for $\eta$: The lhs of (\ref{thresh}) is a function that
starts of at $0$ for $\eta=0$, increases linearly, reaches a maximum of
$1/2\tih$ at $\eta=\tih$, and decays as $1/\eta$ for large $\eta$.
Condition (\ref{thresh}) then limits the integration range of $\eta$ to an
interval $\eta_1\le\eta\le\eta_2$ with 
\begin{equation} \label{eta12}
\eta_{1,2}=\frac{1\pm\sqrt{1-4\delta^2\tih^2}}{2\delta}\equiv \eta_{1,2}(\kappa),
\end{equation}
where  $\delta\equiv c\Delta
\omega/(v_s\omega_c)=c\kappa_x/\omega_c=c\kappa/(\omega_c\Delta
r)=\chi\kappa$, 
$\kappa\equiv \kappa_x\Delta r$,  and $\chi= c/(\omega_c
\Delta r)\ll 1$ (see introduction and Fig.\ref{fig:eta12}).  %
A finite real integration range exists only for $\delta <1/(2\tih)$,
equivalent to $\kappa<\frac{\omega_c\Delta r}{2c\tih}\equiv
\kappa_{\rm Max}$. For given $\eta$, we have
$\kappa\le \kappa_{\rm max}(\eta)\equiv \frac{\omega_c\Delta
  r}{c}\frac{\eta}{\eta^2+\tih^2}\le \kappa_{\rm Max}$. A finite
minimal value of $\kappa$
can be deduced from a maximum desired snapshot size in $x-$direction.
Also the 
requirement $\tau<h/v_s$ may bound the relevant values of $\kappa$
from below, as it leads to a smallest resolvable frequency, and thus
also smallest resolvable $\Delta \omega$: $\Delta\omega>
2\pi/\tau\Rightarrow \kappa=\Delta r \Delta\omega/v_s>\Delta
r\frac{2\pi}{\tau v_s}\equiv \kappa_{\rm min}$.\\ 

As the contributions from areas outside the allowed range
$\eta_1\le\eta\le \eta_2$ (or, correspondingly for negative $\eta$,
$-\eta_2\le\eta\le -\eta_1$) are  
exponentially suppressed, we can limit the integration range of $\eta$ to that
interval for a given $\kappa_x$, and replace the integral
kernel by its approximate form, eq.(\ref{Iasy}), extended to $\beta<0$ by
(\ref{sym}), yielding
\begin{equation}
  \label{eq:Iab1gen}
  K(\alpha,\beta)\simeq
  \sqrt{\frac{2\pi}{|\beta|}}e^{\sign(\beta)i\frac{\pi}{4}}e^{-i\beta} 
\end{equation}
in the allowed range, and zero elsewhere. 
 After the substitution 
$\zeta=\eta/\sqrt{\eta^2+\tih^2}$, the result for $C_{ii}$ can be
written as 
\begin{eqnarray}
C_{ii}(\br_1,\br_1+\Delta r \hat{e}_y,\kappa,\tilde{k}_c)&\simeq\sqrt{2\pi}&\frac{K_6
  e^{-i\kappa \tix_0}}{\sqrt{|\tilde{k}_c|}}\int_{-\infty}^\infty 
d\zeta F(\kappa,\zeta,\tilde{k}_c)e^{-i\tilde{k}_c\zeta}\label{eq:cii}\\
F(\kappa,\zeta,\tilde{k}_c)&=&\left(e^{i\sign(\tilde{k}_c)\pi/4}w(\zeta_1(\kappa),\zeta_2(\kappa),\zeta)
  +e^{-i\sign(\tilde{k}_c)\pi/4}w(-\zeta_2(\kappa),-\zeta_1(\kappa),\zeta)\right)\nonumber\\ 
&&\times\tilde{T}\left(\kappa,\frac{\zeta\tih}{\sqrt{1-\zeta^2}}\right)
\frac{1}{\sqrt{|\zeta|}(1-\zeta^2)}\,, \label{eq:Fz}     
\end{eqnarray}
where $\tilde{T}(\kappa,\eta)\equiv\tilde{T}_{0,\eta\Delta
  r}(\kappa_x)$ 
(with
$\kappa=\kappa_x\Delta r$, $\tix_0=x_0/\Delta r$), and %
$w(\zeta_1,\zeta_2,\zeta)$ is a window function equal to one for
$\zeta_1\le\zeta\le\zeta_2$ and zero elsewhere. The window functions
translate in a straight forward fashion the integration range for $\eta$
into an integration range for $\zeta$.  By definition, $\zeta$ ranges
from $-1,\ldots, 1$. So $\zeta_1,\zeta_2$, lie within this
interval, $-1\le\zeta_1,\zeta_2\le1$, and the window functions take
care of restricting the argument 
$\zeta$ of the integrand to the intervals $\pm[\zeta_1,\zeta_2]$. 
We
have replaced $\omega_1,\omega_2$
by the equivalent information $\kappa\equiv\kappa_x\Delta r$ (related
to $\Delta \omega$) and $\tilde{k}_c=\Delta r \omega_c/c$ (related to
$\omega_c$), and consider $i=j$ only.  Given eq.(\ref{eq:cii}) it is
tempting to try to 
recover  $F(\kappa,\zeta,\tilde{k}_c)$ by Fourier transform. However,
the $\sign(\tilde{k}_c)$ 
functions that appear in $F(\kappa,\zeta,\tilde{k}_c)$ prevent (\ref{eq:cii})
from being a 
simple Fourier integral.  Moreover, from the measured data we only have
$C_{ij}^F$, the filtered version of $C_{ij}$, that is restricted to a
frequency range $\omega_1,\omega_2\in \pm[\omega_0-\pi B,\omega_0+\pi
B]$, where 
$B$ is the bandwidth (20MHz in SMOS for the L-band). We assume here
for simplicity a Gaussian filter and the same for 
both antennas. For a real filter response function $A(t)$, its Fourier
transform must satisfy $A(\omega)=A^*(-\omega)$.  Taking also
$A(\omega)$ as real, we can write it as 
\begin{equation}
  \label{eq:A}
  A(\omega)=(G(\omega;-\omega_0,b)+G(\omega;\omega_0,b))\sqrt{b}\pi^{1/4}\,,
\end{equation}
where
$G(\omega;\omega_0,b)=\exp(-(\omega-\omega_0)^2/(2b^2))/(\sqrt{2\pi}b)$
is a normalized Gaussian 
centered at $\omega_0$ with standard  deviation $b\equiv 2\pi
B$. The factor 
$\sqrt{b}\pi^{1/4}$ assures that for $\omega_0\gg b$, $A(\omega)$ is
normalized according to
$\int_{-\infty}^\infty|A(\omega)|^2\,d\omega=1$. Under the same
condition we have 
\begin{eqnarray}
  \label{eq:CG}
  C_{ii}^F(\br_1,\br_2,\kappa,\tilde{k}_c)&=&
  C_{ii}(\br_1,\br_2,\kappa,\tilde{k}_c)A(\omega_1)A^*(\omega_2)\\
A(\omega_1)A^*(\omega_2)&=&\frac{1}{2}
\left(G(\tilde{k}_c;\tilde{k}_{c0},\frac{\tib}{\sqrt{2}})+ 
      G(\tilde{k}_c;-\tilde{k}_{c0},\frac{\tib}{\sqrt{2}})\right)\,, \label{eq:CG2}   
\end{eqnarray} 
where $\tik_{c0}=\Delta r \omega_0/c=1/\chi\simeq 2932.55$, and
$\tib=\Delta r b/c \simeq 41.89$.
This contains the approximation of using only the ``diagonal'' terms
in the product of $A(\omega_1)A^*(\omega_2)$, i.e.~the ones with
$\sign(\omega_1)=\sign(\omega_2)$, 
 which is justified by the fact %
that $\Delta \omega\ll b\ll\omega_0$.
Fourier transforming $C_{ii}^F$
 with respect to $\tik_c$ (denoted by ${\cal F}_{\tik_c\to \zeta}$)  
 gives a convolution product between the FT of the Gaussians (which is
 $(\sqrt{2}/\tib)G(\zeta; 0, \sqrt{2}/\tib)e^{\pm i\tik_{c0} \zeta}$) and
 $F(\kappa,\zeta,\tilde{k}_c)$, 
 and leads to 
 \begin{eqnarray}
   \label{eq:FC}
   {\cal
     F}_{\tik_c\to \zeta}\left(C_{ii}^F(\br_1,\br_1+\Delta r \hat{e}_y,\kappa,\tilde{k}_c)\sqrt{\frac{|\tik_c|}{2\pi}}
     \frac{e^{i\kappa_x
         x_0}}{K_6}\right)&=&\frac{\sqrt{2}}{\tib}\sum_{\sigma=\pm}\left(G(\zeta;0,\frac{\sqrt{2}}{\tib})\cos(\tik_{c_0}\zeta+\sigma\frac{\pi}{4})\right)\\  
&&\star\left[w(\zeta_1(\kappa),\zeta_2(\kappa),\sigma\zeta)\tilde{T}(\kappa,\frac{\zeta\tih}{\sqrt{1-\zeta^2}})\frac{1}{\sqrt{|\zeta|}(1-\zeta^2)}\right]\nonumber
\,, 
 \end{eqnarray} 
where we used that the sign of $\tik_{c0}$ in \eqref{eq:CG} determines
the one of $\tik_c$ in \eqref{eq:Fz}. %
Thus, we get back the original function
$\tilde{T}(\kappa,\frac{\zeta\tih}{\sqrt{1-\zeta^2}})=\tilde{T}(\kappa,\eta)$,
cut by the two window 
functions and multiplied with $1/(\sqrt{\zeta}(1-\zeta^2))$, 
convoluted with the product of a Gaussian of width $\sqrt{2}/\tib$ and
a rapidly oscillating cosine function.   The factor
$1/(\sqrt{\zeta}(1-\zeta^2))$ can be tracked 
back to the change of variables from $\eta$ to $\zeta$ and will
distort the image at the nadir and at infinity.
Sources at positive or 
negative $\eta$ contribute differently due to the different sign of
the $\pi/4$ phase shift.  This arises already in 
(\ref{eq:Iab1gen}) due to the different phase shift in the asymptotics
of the Struve
functions for negative or positive arguments and leads to the sum over
$\sigma=\pm$. In general,
an exact 
inversion can not be simply done by Fourier transform but needs a
numerical approach.  Nevertheless, we can arrive at an estimation of
the resolution by considering a single point source, as then only one
of the two terms in the sum over $\sigma$ in (\ref{eq:FC})
contributes, and the factor $1/(\sqrt{\zeta}(1-\zeta^2))$ becomes a
simple numerical factor given by the position of the source.   

\subsection{Single point source and geometric resolution}\label{sec.sps}
\subsubsection{Correlation function and reconstructed image}
Let the point source be at position $x''=0,y''=\eta_s \Delta r$ and with
polarization $i$, where
$\eta_s$ is situated in the allowed range $0\le\eta_1(\kappa)\le
\eta_s\le \eta_2(\kappa)$ 
for some $\kappa$ in the desired range up to the largest considered
$\kappa=2\pi/p_x$, where 
$p_x$ is the pixel size.  We 
thus have 
\begin{equation} \label{IPS}
T(\br'')=T_0\delta(x'')\delta(y''-\eta_s\Delta r)\Delta r^2\,,
\end{equation}
which together with eq.(\ref{Ix0}) yields
\begin{equation} \label{IPS2}
\tilde{T}(\kappa,\eta)=\frac{T_0\Delta r}{\sqrt{2\pi}}\delta(\eta-\eta_s).
\end{equation}
As $\eta$ is in the allowed range, we can use the approximate analytical
form of the integral kernel, eq.(\ref{eq:Iab1gen}), to get from
eq.(\ref{eq:cii}) the 
correlation function 
\begin{eqnarray} \label{Cijp} 
C_{ii}(\br_1,\br_1+\Delta r
\hat{e}_y,\kappa,\tilde{k}_c)&=&K_6T_0\Delta r 
\frac{e^{-i\kappa\tilde{x}_0}e^{i\sign(\tik_c)\pi/4}}{\sqrt{|\tik_c|\eta_s}(\eta_s^2+\tih^2)^{1/4}}e^{-i\tik_c\frac{\eta_s}{\sqrt{\eta_s^2+\tih^2}}} \theta(\kappa_{\rm
  max}(\eta_s)-|\kappa|)\,,
\end{eqnarray}
where $\theta(x)$ is the Heaviside theta-function. 
Considering (\ref{eq:FC}), we may define an approximative reconstructed source function
suitable for sources at $\eta_s>0$ through 
 \begin{equation}
   \label{eq:frec}
\tilde{T}_{\rm rec}(\kappa,\eta) \equiv  {\cal N}{\cal
     F}_{\tik_c\to\zeta}\left(C_{ii}^F(\br_1,\br_1+\Delta r \hat{e}_y,\kappa,\tilde{k}_c)\sqrt{\frac{|\tik_c|}{2\pi}}\frac{e^{i\kappa_x  
       x_0}}{K_6} \right)\,,
 \end{equation}
where ${\cal N}$ is a normalization constant. Due to the
$\kappa-$dependence of the window functions, and the 
$\zeta$ dependence of the integral transform as compared to a simple
Fourier-transform of $\tilde{T}$, one cannot get a
normalization constant independent of the source field. In particular,
for the single point 
source, ${\cal N}$ would depend on the position of the point
source. However, we use $T_{\rm rec}$ only for estimating the
geometric and radiometric resolution.  For the former, all prefactors
are irrelevant.  For the latter, we avoid the problem by calculating
relative uncertainties of $\sigma(T_{\rm rec})/T_{\rm rec}$ only,
where any prefactor cancels. We hence set ${\cal N}=1$ in the
following.   \\ 
 
Inverting the Fourier transform in $\kappa$ leads to 
\begin{equation}
  \label{eq:Trec}
  T_{\rm rec}(x,y)=\frac{1}{2\pi K_6 \Delta r}\int_{-\infty}^\infty
  d\kappa\int_{-\infty}^\infty d\tik_c
  e^{-i\kappa(\tix-\tix_0)}e^{i\tik_c\zeta}C_{ii}^F(\br_1,\br_1+\Delta r \hat{e}_y,\kappa,\tik_c)\sqrt{\frac{|\tik_c|}{2\pi}}\,. 
\end{equation}
This equation is valid for all sources located in the positive $y$
plane, not necessarily point sources. 
When we re-express the correlation function through \eqref{eq:cii} and
perform the Gaussian integral over $\tik_c$, we find a direct
approximate formal 
relation between the FT of the original $T(x,y)$ in the upper half
plane, and its 
reconstructed image $T_{\rm 
  rec}(x,y)$, 
\begin{align}
  T_{\rm rec}(x,y)=\frac{1}{2\pi\Delta r}\int_{-\infty}^\infty
  d\kappa\int_{-\infty}^\infty d\zeta'
  \frac{\tilde{T}(\kappa,\frac{\zeta'\tih}{\sqrt{1-\zeta'^2}})}{\sqrt{|\zeta'|(1-\zeta'^2)}}e^{-i\kappa
  \tix}e^{-\tib^2(\zeta-\zeta')^2/4}\nonumber\\
\times\left(
w(\zeta_1(\kappa),\zeta_2(\kappa),\zeta')\cos(k_{c0}(\zeta-\zeta')+\frac{\pi}{4})
+
w(\zeta_1(\kappa),\zeta_2(\kappa),-\zeta')\cos(k_{c0}(\zeta-\zeta')-\frac{\pi}{4}) 
\right)\,.\label{eq:TtiToT}
\end{align}
Using this expression, or by inserting (\ref{Cijp}) into
(\ref{eq:CG}), and the resulting filtered 
correlation function into (\ref{eq:Trec}), we find the reconstructed
image of the single point source 
\begin{eqnarray}
  \label{eq:Trecxy}
  T_{\rm
    rec}(x,y)&=&\frac{T_0\sqrt{\zeta_s}(1-\zeta_s^2)}{\sqrt{2}\pi^{3/2}\chi
  \tih^2}e^{-(\zeta-\zeta_s)^2\tib^2/4}\cos(\tik_{c0}(\zeta-\zeta_s)+\frac{\pi}{4})
\sinc(\kappa_{\rm max}(\zeta_s)\tix/\pi)\,,
\end{eqnarray}
where $\zeta_s=\eta_s/\sqrt{\eta_s^2+\tih^2}$, $\kappa_{\rm max}(\zeta_s)\equiv
\zeta_s\sqrt{1-\zeta_s^2}/(\chi\tih)$,  and $\sinc(x)\equiv \sin(\pi
x)/(\pi x)$. 

\subsubsection{Geometric resolution}
We see that the
reconstructed image of the point source is a series of narrow
peaks spaced by the inverse of $\tik_{c_0}$ due to the rapidly
oscillating $\cos$-function, under an approximate Gaussian in
$y$-direction centered at the position of the source
with a width in $\eta$ given by $\Delta
\eta=\sqrt{2}\sqrt{\eta_s^2+\tih^2}/\tib\ge 
\sqrt{2}\tih/\tib=hc/((\Delta r)^2\sqrt{2}\pi B)$.  
It reminds one of a
diffraction 
image from a double slit, even though there the envelope is a
$\sinc$-function, not a Gaussian. Nevertheless, we adapt the
definition of resolution from that example, namely that the best
resolution is obtained from the smallest shift that makes a peak move
into the next trough.  This leads to
\begin{equation}
  \label{eq:resy}
  \tik_{c_0}\frac{\partial}{\partial
    \eta}\frac{\eta}{\sqrt{\eta^2+\tih^2}}\left|_{\eta=\eta_s}\right.\Delta \eta\simeq \pi\,,
\end{equation}
hence $\Delta y=\Delta r\pi (\eta_s^2+\tih^2)^{3/2}/(\tik_{c_0}\tih^2)$. 
For $y \simeq h$, this is of the 
order  $2\sqrt{2}\pi hc/(\Delta r \omega_c)=\chi h$. 
The numerical value for the
standard parameters gives $\Delta y\simeq$ 2.1\,km, i.e.~a resolution
of the order of a kilo-meter.
However, for actually achieving this resolution for an extended
source, one has to face two issues: {\it i.}) The reconstructed point-source
image should 
be brought as close as possible to a single narrow peak; and {\it
  ii.}) one has to deal with the different phases from sources at
positive or negative $\eta$.  The first issue can be addressed by
superposing correlation functions from pairs of antennas at different
separation, and/or changing the considered central frequency. This 
shifts the pattern of peaks due to the cos-function, and one can
engineer a rather narrow central peak (see \cite{NumPaper} for
details).   The second issue should be absent in a numerically exact
inversion of the integral kernel.
The Gaussian envelope has a width $hc/(\sqrt{2}\pi B \Delta r)$ given
by the inverse bandwidth, which is much larger than the width of a
single peak, namely 
by a factor $\omega_c/(4\pi B)\simeq 35$ for the standard parameters. \\

The resolution in $x$-direction follows from the effective wave vector
$\kappa_{\rm max}$ in the $\sinc$ function. It depends on
the position of the source and reaches its maximum possible value
$\kappa_{\rm Max}$ for $\eta_s=\tih$ (i.e.~$y_s=h$). The inverse of
$\kappa_{\rm Max}$ thus gives the best possible resolution in $x$-direction:
\begin{equation}
  \label{eq:dx}
  \Delta x\ge \frac{\Delta r}{\kappa_{\rm Max}}=\frac{2hc}{\omega_c
    \Delta r}\,.
\end{equation}
We conclude that both in $x$- and $y$-direction one can expect a
geometric resolution of the order $h \chi=c/(\Delta r\omega_c)$ for
sources close to $y=h$.  
For sources close to $y=0$, $\kappa(\eta_s)$
goes to zero $\propto \eta_s$, whereas for larger $y_s$ the
decay of $\kappa(\eta_s)$ is $\propto 1/\eta_s$. The geometric
resolution in $x$-direction deteriorates correspondingly. The
resolution in $y$-direction, on the other hand, depends only weakly 
on the source-position, as
$(\eta_s^2+\tih^2)^{3/2}/(\tik_c\tih^2)$ increases monotonically
from $\tih$ at $y_s=0$ to $2\sqrt{2}$  at $y_s=h$, and keeps growing
slowly beyond $y_s=h$. 
It is remarkable that correlating electric fields at two different
frequencies can lead 
to a resolution that is given by the central frequency. \\

The definition of $\kappa_{\rm  max}$ is based on the
request that the stationary phase approximation (SPA)
holds in the regime $\beta\gg \alpha$.   In practice, the SPA is
almost always better than expected, 
such that in the end the result $h\chi $  might be a conservative
estimate of the geometric resolution.

\section{Radiometric resolution}
Besides the geometric resolution, the radiometric resolution (RR),
i.e. the smallest  
difference in temperature that the system can measure for a given
pixel, is the most important characteristics of the satellite imaging
system. Here we calculate the RR for the idealized situations of a
single point source considered above and for a uniform temperature
field  in the positive half-plane $y>0$.

\subsection{Fluctuations of the reconstructed temperature profiles}  
The idea behind the calculation of RR is that
the electric field measurements yield random values, whose fluctuations  and
correlations reflect the thermal nature of the radiation field. Thus, if
with the same field $T(x,y)$ one repeated the measurements many times, one
would obtain different correlation functions in each run, and thus,
after inverting the linear relationship between $C_{ij}$ and $T(x,y)$, also
different reconstructed $T(x,y)$ (called $T_\text{rec}$ in the
following) in each run. The (relative) RR is then defined as the 
standard deviation $\sigma(T_\text{rec}(x,y))$ divided by the average
$T_\text{rec}(x,y)$ for a 
given position $x,y$. In general it will vary as function of $x,y$, and also
depend on the temperatures at all positions, a behavior well known from
standard spatial aperture synthesis. In reality, things become still a bit
more complicated, as the measured signal is a superposition of the
e.m.~field emitted by the antenna itself (at temperature $T_a$), and the
radiated field from the surface of Earth.  However, these fields are
uncorrelated, and their averaged squares just add up. For
simplicity, we will neglect the noise contribution of the antennas in this
first analysis, which amounts to calculating lower bounds of
$\sigma(T_\text{rec})$. 

Starting point of the calculation is the
assumption that the current fluctuations $\bj(\br'',t)$ which are at the
origin of the radiated thermal field are described by a random
Gaussian process, both in 
time and space (see sec.~\ref{sec.model}). This implies
immediately, that also the temporal FT 
$\tilde{\bj}(\br'',\omega')$ of the 
current fluctuations is a Gaussian process, now over space and frequency.
Finally, the connection between $\tilde{\bj}(\br'',\omega')$ and
$\tilde{E}_{\br_1}(\omega_1)$  is linear, which implies that
$\tilde{E}_{\br_1}(\omega_1)$ is a Gaussian process over $\br_1$ and
$\omega_1$. By the nature of this variable, it is a complex Gaussian
process. One easily shows that the
average of $\tilde{E}_{\br_1}(\omega_1)$ equals zero (if the average of all
current components is, which must be true at  thermal equilibrium).  The
correlation function $C_{ij}$ is the (complex) 
covariance matrix of this Gaussian process, and all higher correlations can
be expressed in terms of it. \\

In order to assess the fluctuations of $T_{\rm rec}$ we first define a product of
Fourier coefficients of $\bE$ from a single run (denoted by a $\hat{}$
), 
\begin{eqnarray}
  \label{eq:Chat}
 \hat{C}(\br_1,\br_2,\kappa,\tik_c)\equiv \hat{C}_{zz}(\br_1,\br_2,\omega_1,\omega_2)&\equiv&
  \hat{\tilde{E}}_{z,\br_1}(\omega_1)\hat{\tilde{E}}_{z,\br_2}^*(\omega_2) \\
&=&\frac{1}{2\pi}\int dt_1\int dt_2
\hat{E}_{z,\br_1}(t_1)\hat{E}_{z,\br_2}(t_2)e^{-i\omega_1 t_1+i \omega_2 t_2}\,,
\end{eqnarray}
and its corresponding filtered version
$\hat{C}^F(\br_1,\br_2,\kappa,\tik_c)=\hat{C}(\br_1,\br_2,\kappa,\tik_c)A(\omega_1)A^*(\omega_2)$
(with $\omega_{1},\omega_2$ expressed in terms of $\kappa,\tik_c$). 

The fluctuations of $T_{\rm rec}(x,y)$ are defined as $\Delta T_{\rm
  rec}(x,y)\equiv \langle T_{\rm rec}(x,y)^2\rangle-\langle
T_{\rm rec}(x,y)\rangle^2$, where the average is over the thermal
ensemble.
With
$T_{\rm rec}(x,y)$ from (\ref{eq:Trec}), one finds 
\begin{eqnarray}
  \label{eq:dTrec}
  \Delta T_{\rm rec}(x,y)&=& \frac{1}{K_6^2(2\pi)^3\Delta r^2}\int
  d\kappa_1\,d\kappa_2\,d\tik_{c1}\,d\tik_{c2}\sqrt{|\tik_{c1}\tik_{c2}|}e^{-i(\kappa_1-\kappa_2)(\tix-\tix_0)}e^{i(\tik_{c1}-\tik_{c2})\zeta}\nonumber\\
&&\times\Big(\langle\hat{C}^F(\br_1,\br_2,\kappa_1,\tik_{c1})\hat{C}^{F*}(\br_1,\br_2,\kappa_2,\tik_{c2})\rangle\nonumber\\
&&-\langle\hat{C}^F(\br_1,\br_2,\kappa_1,\tik_{c1})\rangle\langle\hat{C}^{F*}(\br_1,\br_2,\kappa_2,\tik_{c2})\rangle\Big)
\,. 
\end{eqnarray}
It can be shown (see Sec.\ref{sec.circular}) that in the narrow
frequency intervals considered 
here the Gaussian random processes given by the
$\tilde{E}_{z,\br_{i}}(\omega)$ 
is circularly symmetric.  It hence enjoys the property
(see eq.~8.250 in \cite{Barrett03}), 
\begin{equation}
  \label{eq:umom}
  \langle E_i E_j E_k^* E_l^*\rangle = \langle E_i E_k^*\rangle\langle E_j
  E_l^*\rangle +\langle E_i E_l^*\rangle\langle E_j
  E_k^*\rangle\,,
\end{equation}
where we have abbreviated
$E_i\equiv\tilde{E}_{z,\br_{(i\, {\rm mod} \,2)}}(\omega_i)$.  
The correlation function contained in the large parentheses of the
second line of \eqref{eq:dTrec} becomes
\begin{equation}
  \label{eq:1324}
  C_{zz}^F(\br_1,\br_1,\kappa_{13},\tik_{c13})  C_{zz}^{F*}(\br_2,\br_2,\kappa_{24},\tik_{c24})\,,
\end{equation}
with $\kappa_{ij}=\Delta r(\omega_j-\omega_i)/v_s$,
$\tik_{cij}=\Delta r(\omega_i+\omega_j)/(2c)$ $\forall i,j$, and where
we have used $\langle\hat{C}^F(\br_1,\br_2,\kappa,\tik_c)
\rangle=C^F(\br_1,\br_2,\kappa,\tik_c)$. 

The fact that $C_{zz}^F$ and $C_{zz}^{F*}$ in \eqref{eq:1324} contain the same position
arguments twice makes that we cannot evaluate it directly through
eq.\eqref{Cfin}, as the coordinate transformation to dimensionless
variables based on the rescaling with $\Delta r$ becomes singular.  We
therefore have to go back a step to eq.\eqref{C3} which yields
\begin{eqnarray}
  \label{eq:C3rr}
  C_{zz}(\br_1,\br_1,\kappa,\tik)&=&K_6e^{-i\kappa_x x_0}\int_{-\infty}^\infty\int_{-\infty}^\infty
  dx'\,dy'\frac{\tilde{T}_{0,y'}(\kappa_x)e^{-i\kappa_x'x'}}{x'^2+y'^2+h^2}\\
&=&K_6e^{-i\kappa_x x_0}\pi \int dy' \frac{\tilde{T}_{0,y'}(\kappa_x)e^{-|\kappa_x|\sqrt{y'^2+h^2}}}{\sqrt{y'^2+h^2}}\,.\label{eq:C3rrb}
\end{eqnarray}
Comparing with \eqref{Cfin} we see that this result corresponds
formally to $\omega_c=0$ in that equation, rather than $\Delta r=0$,
and \eqref{eq:C3rrb} is recovered by using the exact result \eqref{b0}. 
We can now re-introduce dimensionless variables via the same
rescaling with $\Delta r$, where, however, 
$\Delta r$ is still given by $\Delta r=|\br_2-\br_1|$, and $\br_i$
denote as before the positions of the two antennae at $t=0$, only one
of which still enters as argument in $C_{zz}(\br_1,\br_1,\kappa_{13},\tik_{13})$,
respectively $C_{zz}^{F*}(\br_2,\br_2,\kappa_{24},\tik_{c24})$.  This
gives
\begin{eqnarray}
  \label{eq:C3rr2}
  C_{zz}(\br_1,\br_1,\kappa,\tik)&=&K_6e^{-i\kappa \tix_0}\pi 
  \int d\eta'\frac{\tilde{T}_{0,\eta'}(\kappa)e^{-|\kappa |\sqrt{\eta'^2+\tih^2}}}{\sqrt{\eta'^2+\tih^2}}\,.
\end{eqnarray}
We calculate $\Delta T_{\rm rec}$ for the single
point source considered in Sec.~\ref{sec.sps} at the
position of the source, i.e.~$\Delta T_{\rm
  rec}(x_s,y_s)=\langle T_{\rm rec}(x_s,y_s)^2\rangle-\langle
T_{\rm rec}(x_s,y_s)\rangle^2$, and for a uniform temperature field. 

\subsection{Single point source}
For the single point source at position $(0,y_s)$, the correlation
function $C_{zz}(\br_1,\br_1,\kappa,\tik)$ becomes (see \eqref{IPS2})
\begin{equation}
  \label{eq:CrrSPS}
  C_{zz}(\br_1,\br_1,\kappa,\tik)=K_6\sqrt{\frac{\pi}{2}}T_0
  e^{-i\kappa \tix_0}\frac{e^{-|\kappa|\sqrt{\eta_s^2+\tih^2}}}{\sqrt{\eta_s^2+\tih^2}}\,.
\end{equation}
Insert this into eq.\eqref{eq:dTrec} to find
\begin{eqnarray}
  \label{eq:dTrec0ys}
  \Delta
  T_{\rm rec}(0,y_s)&=&\frac{T_0^2}{16\pi^2}\int
  d\kappa_{12}d\kappa_{34}d\tik_{c12}d\tik_{34}
\sqrt{\left|\tik_{c12}\tik_{c34}\right|}\frac{e^{-(|\kappa_{13}|+|\kappa_{24}|)\sqrt{\eta_s^2+\tih^2}}}{\eta_s^2+\tih^2}\nonumber\\ 
&&\times\left(
G(\tik_{c13};\tik_{c0},\frac{\tib}{\sqrt{2}})+G(\tik_{c13};-\tik_{c0},\frac{\tib}{\sqrt{2}})  
\right)
\left(
G(\tik_{c24};\tik_{c0},\frac{\tib}{\sqrt{2}})+G(\tik_{c24};-\tik_{c0},\frac{\tib}{\sqrt{2}})
 \right)\nonumber\\
&&\times e^{i(\kappa_{12}-\kappa_{34}-\kappa_{13}+\kappa_{24})\tix_0+i(\tik_{c12}-\tik_{c34})\zeta_s}\,.
\end{eqnarray}  
We change integration variables to
$\kappa_{13},\tik_{c13},\kappa_{24},\tik_{c24}$. The
Jacobian is 1. In 
the product of the Gaussians only the diagonal terms (i.e.~with the
same signs in front %
of $\tik_{c0}$) contribute in the relevant regime
$\tik_{c0}\gg \tib$, as for opposite signs
$\tik_{c12}\simeq\tik_{c34}\simeq 0$. Finally, we approximate
\begin{equation}
\tik_{c12}=\tik_{c34}\simeq \tik_{c0}  \label{app}
\end{equation}
 and pull that factor out from
the integral which is
permissible for all ranges of variables for which the product of
Gaussians is non-negligible. The integrals can then be performed, and
we find for the standard deviation $\sigma(T_{\rm rec}(0,y_s))\equiv
\sqrt{\Delta(T_{\rm rec}(0,y_s))}$ 
\begin{equation}
   \label{eq:sTfin}
   \sigma(T_{\rm rec}(0,y_s))=\frac{T_0}{\sqrt{2}\pi}\frac{|\tik_{c0}|^{1/2}}{\eta_s^2+\tih^2+\frac{\beta^2\zeta_s^2}{4}}\simeq\frac{T_0}{\sqrt{2}\pi}\frac{|\tik_{c0}|^{1/2}}{\eta_s^2+\tih^2}\,,
\end{equation}
where in the last step we have used that $\tih\gg 1$ and $\beta
\zeta_s\ll 1$ (where once more $\beta=v_s/c$). \\
From \eqref{eq:Trecxy} we find
\begin{equation}
  \label{eq:T0ys}
  T_{\rm rec}(0,y_s)=\frac{T_0\sqrt{\eta_s}}{2\pi^{3/2}\chi (\eta_s^2+\tih^2)^{5/4}}\,.
\end{equation}
Combined with \eqref{eq:sTfin} we obtain the relative RR
\begin{equation}
  \label{eq:sTrelSPS}
  \frac{\sigma(T_{\rm rec}(0,y_s))}{T_{\rm rec}(0,y_s)}=\sqrt{\frac{2\pi\chi}{\eta_S}}(\eta_s^2+\tih^2)^{1/4}\,.
\end{equation}
For $\eta_s=\tih$, the relative RR is of order 0.055, corresponding at
$T=300$\,K to $\sigma(T_{\rm rec}(0,h))\simeq 16.5$\,K.

\subsection{Uniform temperature field}
We now look at the
second standard situation considered commonly for the determination of
the radiometric resolution, namely a field of constant
temperature. More precisely, we consider
\begin{equation}
  \label{eq:Tuni}
  T(x,y)=\left\{
    \begin{array}{cc}
      T_0&0\le y \le \hat{y}\\
0 & \mbox{else}
    \end{array}
\right.\,.
\end{equation}
The restriction to sources in the upper plane is due to the fact that
we still want to use eq.~\eqref{eq:Trec} for calculating the
reconstructed temperature profile.  The cut-off $\hat{y}$ arises
physically from the size of the Earth and prevents a divergence of the
correlation function. \\
From \eqref{Ii} we obtain 
\begin{equation}
  \label{eq:Ttiluni}
  \tilde{T}(\kappa_x,\eta)=\left\{
    \begin{array}{cc}
      \sqrt{2\pi}\delta(\kappa_x)T_0&0\le y \le \hat{y}\\
0 & \mbox{else}
    \end{array}
\right.\,.
\end{equation}
 
The correlation function \eqref{eq:C3rr2} becomes 
\begin{eqnarray}
  \label{eq:Cuni}
  C_{ii}(\br_1,\br_1,\kappa,\tik_c)&=&K_6\sqrt{2}\pi^{3/2}T_0e^{-i\kappa\tix_0}\delta(\kappa)\Delta
  r \int_{0}^{\hat{\eta}}\frac{d\eta}{\sqrt{\eta^2+\tilde{h}^2}}\\
&=&K_6\sqrt{2}\pi^{3/2}T_0e^{-i\kappa\tix_0}\delta(\kappa)\Delta r \ln\left(\frac{\hat{\eta}+\sqrt{\hat{\eta}^2+\tih^2}}{\tih}\right)\,.
\end{eqnarray}
with $\hat{\eta}\equiv 
\hat{y}/\Delta r$. %
For $\hat{y}=R_E\simeq 6370$\,km the radius of
Earth, and $\Delta r=100$\,m, $\hat{\eta}=63700$. 
We see that here the correlation function is perfectly diagonal in
frequency, which of course reflects the lack of structure of the
temperature field in $x$-direction.  Hence, we can set everywhere 
$\kappa=0$, which greatly simplifies the analysis. 
The cutoff $\hat{\eta}$ in eq.(\ref{eq:Cuni}) 
prevents a logarithmic divergence
that arises from 
$1/\sqrt{\eta^2+\tih^2}\sim 1/\eta$ for
$\eta\to\infty$. Eq.\eqref{eq:Cuni} can be extended  
to a temperature field that is uniform everywhere from $-\hat{y}$ to
$\hat{y}$. In this case, the $\zeta$-integral starts at $-1+\epsilon$
rather than at 0. However, in this situation we cannot use
\eqref{eq:Trec} anymore as it is valid only for sources at positive
$y$ (see the discussion after \eqref{eq:FC}).\\

Eq.(\ref{eq:Cuni}), when inserted into \eqref{eq:dTrec}, and with the
same change of integration variables and approximation \eqref{app},  leads to
\begin{equation}
  \label{eq:DtUni}
  \sigma(T_{\rm
    rec}(0,y))=\frac{T_0}{\sqrt{2}}\sqrt{|\tik_{c0}|}\ln\left(\frac{\hat{\eta}+\sqrt{\hat{\eta}^2+\tih^2}}{\tih}\right)\,. 
\end{equation}

The reconstructed temperature field
\eqref{eq:TtiToT} is given by
\begin{equation}
  T_{\rm rec}(x,y)=
\frac{T_0}{\sqrt{2\pi}}\int_{0}^{\hat{\zeta}}d\zeta'\frac{e^{-\tib^2(\zeta-\zeta')^2/4}}{\sqrt{|\zeta'|}(1-\zeta'^2)}\cos(\tik_{c0}(\zeta-\zeta')+\frac{\pi}{4})\,.
\label{eq:TrecUni} 
\end{equation}
Unfortunately, no closed analytical form could be found for the
remaining integral, and even a numerical evaluation is not straight
forward, as the Gaussian yields a very narrow peak, broader, however,
than the period of the $\cos$-function. But we can get an
estimate of $T_{\rm rec}$ by replacing the Gaussian (normalized to an
integral equal 1) with a rectangular
peak of width $a\sigma$ and height $1/(a\sigma)$ centered, as the
Gaussian, at $\zeta$.  Here, $\sigma=\sqrt{2}/\tib$, and $a$ is a
parameter of order 1. This gives
\begin{equation}
  T_{\rm rec}(x,y)=
\frac{T_0}{a}\int_{\max(0,\zeta-a\sigma/2)}^{\min(\hat{\zeta},\zeta+a\sigma/2)}d\zeta'\frac{\cos(\tik_{c0}(\zeta-\zeta')+\frac{\pi}{4})}{\sqrt{|\zeta'|}(1-\zeta'^2)}\,.  
\label{eq:TrecUniBox} 
\end{equation}
A numerical evaluation of the integral is now relatively straight
forward and shows a slowly varying $T_{\rm rec}(0,y)$ as function of
$\zeta$ in the interval $\zeta\in [a\sigma/2,\hat{\zeta}-a\sigma/2 ]$,
whereas outside this interval it oscillates rapidly. The slow
variation arises from the factor $\sqrt{|\zeta'|}(1-\zeta'^2)$ that
distorts this approximately reconstructed image.  Pulling out this
slowly varying factor
in order to get an analytical estimate of the
order of magnitude of $T_{\rm rec}(x,y)$, we are led to
\begin{equation}
  \label{eq:TrecUniBox2}
  T_{\rm rec}(x,y)\simeq T_0\frac{\sqrt{2}}{a\tik_{c0}}\frac{1}{\sqrt{|\zeta|}(1-\zeta^2)}\sin\left(\frac{a\tik_{c0}}{\sqrt{2}\tib}
\right)
\end{equation}
for $\zeta\in [a\sigma/2,\hat{\zeta}-a\sigma/2 ]$. Hence, in this
interval and apart from the distorting factor
$1/\sqrt{|\zeta|}(1-\zeta^2)$ identified previously, we recover a
constant temperature field.  The value of the reconstructed 
temperature  
depends on the precise value of $a$  as well as the ratio
$\tik_{c0}/\tib$.  Outside the mentioned interval, $T_{\rm rec}(x,y) $
oscillates again as function of $\zeta$, which can be understood from
the fact that the box is cut-off when $\zeta$ gets within a distance
$a\sigma/2$ of $0$ or $\hat{\zeta}$. The sought-for order of magnitude
can be estimated from the maximum value of \eqref{eq:TrecUniBox2} as
function of $a$.  As for standard parameters $\tik_{c0}/\tib\simeq
70$, we can bound the $\sin$-function by one (while still having
$a\sim 1$), in which case we obtain $T_{\rm rec}(x,y)\simeq
T_0/\tik_{c0}=T_0\chi$ in the mentioned $\zeta$-interval. With all
this, and approximating $\sqrt{\tih^2+\hat{\eta}^2}\simeq \hat{\eta}$
in the logarithm in \eqref{eq:DtUni},  we find the order of magnitude
$\sigma(T_{\rm   rec}(x,y))/T_{\rm  
  rec}(x,y)\sim \ln(2\hat{\eta}/\tih)/\chi^{3/2}$ For $\zeta\sim 1$,
this is of order $\sim 10^5$ for standard parameters,  i.e.~a
catastrophically large uncertainty. A small value of $\sigma(T_{\rm
  rec}(x,y))/T_{\rm rec}(x,y)$ is possible only if
$\sqrt{|\zeta|}(1-\zeta^2)$  is very small, but apart from the fact
that one should not rely on this image-distorting factor, it could only
be sufficiently small for $y$ unrealistically close to the nadir.\\ 

If one traces back the difference to the single-point source, one
realizes that while $\sigma(T_{\rm rec})$ scales in both cases as
$1/\sqrt{\chi}$, the difference comes from $T_{\rm rec}$ itself: for
the single point source it is of order $1/\chi$, but for uniform $T$
in the upper half plane of order $\chi$, which explains a factor
$1/\chi^2$ worse relative RR for the latter compared to the
former. The factor $1/\chi$ in the single-point $T_{\rm rec}$
arises from the cut-off of the $\kappa$ integral: $\kappa_{\rm Max}$
scales as $1/\chi$, and for $x=0$ the $\kappa$-integral in
\eqref{eq:Trec} just gives a factor $2\kappa_{\rm Max}\sim 1/\chi$, as
the correlation-function is independent of $\kappa$ in this case. On
the other hand, for constant temperature in the upper half-plane, the
cut-offs $\kappa_{\rm Max}$ do not play a role, as the
$\delta(\kappa)$-function picks up only $\kappa=0$. This leads to the
loss of one factor $1/\chi$ in $T_{\rm rec}$.  The second one comes
from the integration over $\zeta'$ in \eqref{eq:TrecUni}: the rapidly
oscillating $\cos$-function leads to a factor $1/\tik_{c0}=\chi$,
whereas for the point-source  only a single point $\zeta=\zeta_s$
contributes, such that the cosine is of order one.\\

In the light of the RR of standard radiometers that typically scales as
$\sigma(T)\propto 1/\sqrt{B t_{\rm int}}$, where $t_{\rm int}$ is the
integration time, the fact that $\sigma(T_{\rm rec})$ in
eqs.(\ref{eq:sTfin},\ref{eq:DtUni}) is independent of the bandwidth is
rather surprising. 
Formally, the disappearing of $\tib$ can be traced back to using the
lowest order in the Laplace approximation of \eqref{eq:dTrec0ys}. The
next 
order corrections are of order $\tib$, such that $\sigma(T_{\rm
  rec})\to \sigma(T_{\rm
  rec})(1+{\cal O}(\tib))$.  One expects the sign of the correction
to be positive, as the integrand is positive everywhere, and the
lowest order approximations amounts to replacing the Gaussians by
normalized Dirac-delta functions.  Hence, for small but finite $\tib$, 
$\sigma(T_{\rm rec})$ is expected to increase with $\tib$, which is
contrary to the behavior of standard radiometers.  Standard 
radiometers are based on the van Cittert-Zernike theorem, which gives
the reconstructed temperature field as Fourier transform of the
observed visibilities at a fixed frequency. Different frequencies at
the source are uncorrelated, and the scaling of $\sigma(T_{\rm
  rec})/T_0\propto 1/\sqrt{B t_{\rm int}}$ just reflects averaging over a number
of independent measurements that scales $\propto B \, t_{\rm int}$. In  FouCoIm
the information is in the correlation between different, very narrowly
spaced Fourier components, and averaging over the central frequency
does not lead to additional information (see also
Sec.\ref{sec.freqpairs}). 
Therefore a larger bandwidth does not
improve RR.
\section{Noise reduction}\label{sec.noisered}
The bad signal-to-noise ratio for the radiometric resolution in the
case of a uniform temperature field in the upper half plane, makes 
it essential to consider measures that lead to a noise reduction and
in particular averaging schemes.

\subsection{Averaging over time}
Instead of examining $T_{\rm rec}(x,y)$, we consider here for
simplicity directly the  
fluctuations of the measured ''single shot'' correlation 
function $\hat{C}_{ij}$.  The first idea that comes to mind for
reducing the fluctuations of 
$\hat{C}_{ij}$ is to average over the 
origin of the time-interval from which we construct the
Fourier transform.  Note that this is very different from an ensemble that
one would obtain by displacing the initial position $\br_i$. 
But averaging over the origin of time only leads to an overall factor:
\begin{eqnarray}
  \label{eq:Chat2}
  C^{exp}(\br_1,\br_2,\omega_1,\omega_2)&\equiv&\int_{-\tau_a/2}^{\tau_a/2}dt
\int dt_1\int dt_2
E_{z,\br_1}(t_1+t)E_{z,\br_2}(t_2+t)e^{-i\omega_1 t_1+i \omega_2 t_2}\\
&=&
\sinc((\omega_1-\omega_2)\frac{\tau_a}{2\pi})\hat{C}(\br_1,\br_2,\omega_1,\omega_2) \,.
\end{eqnarray}
So it is obvious that this kind of averaging is useless. This is
indeed to be expected, as all available data were already used. The
situation improves only slightly if the FTs are calculated
from a finite 
stretch of data (say over a duration $\tau_F$). Then shifting the origin in
time will include some new random data, but since we must have $\tau_F\gg
\tau_a$, it is clear that we still use essentially the same data with the
exception of some new data points at the edge of the interval of length
$\tau_F$. 

\subsection{Additional frequency pairs}\label{sec.freqpairs}
Using only a small frequency separation of width $\Delta\omega$ about the
central frequency $\omega_c$ which itself is allowed to vary over a
large bandwidth $B$ appears to be a very wasteful use of all the
pairs of frequency components
$(\tilde{E}_{z,\br_1}(\omega_1),\tilde{E}_{z,\br_2}(\omega_2))$. Can
we use different 
measured correlations $\hat{C}(\br_1,\br_2,\omega_1,\omega_2)$ with
sufficiently different $\omega_c=(\omega_1+\omega_2)/2$ as independent
data for improving the radiometric sensitivity? In order to answer
this question, we need to 
calculate the covariance matrix $V$ between two different correlators,
\begin{equation}
  \label{eq:V}
  V\equiv\langle
  \hat{C}(\br_1,\br_2,\omega_1,\omega_2)\hat{C}^*(\br_1,\br_2,\omega_1',\omega_2')\rangle - \langle
  \hat{C}(\br_1,\br_2,\omega_1,\omega_2)\rangle\langle\hat{C}^*(\br_1,\br_2,\omega_1',\omega_2')\rangle\,,
\end{equation}
as well as the pseudo-covariance matrix $M$,
\begin{equation}
  \label{eq:M}
  M\equiv\langle
  \hat{C}(\br_1,\br_2,\omega_1,\omega_2)\hat{C}(\br_1,\br_2,\omega_1',\omega_2')\rangle - \langle
  \hat{C}(\br_1,\br_2,\omega_1,\omega_2)\rangle\langle\hat{C}(\br_1,\br_2,\omega_1',\omega_2')\rangle\,.
\end{equation}
Both matrices together determine the statistical properties of the
random process $\hat{C}(\br_1,\br_2,\omega_1,\omega_2)$.  Note that
despite the fact that 
$E_{z,\br}(\omega)$ can be considered a circularly symmetric Gaussian process (see 
Appendix) over $\br$ and $\omega$ in the narrow frequency band
we are interested in, the same is not true for 
$\hat{C}(\br_1,\br_2,\omega_1,\omega_2)-\langle
\hat{C}(\br_1,\br_2,\omega_1,\omega_2)\rangle$ (which is not even Gaussian). 
We need
to know whether both $V$ and $M$ essentially vanish for almost 
all pairs of pairs of frequencies, with the first pair
$(\omega_1,\omega_2)$ in a first 
region (notably in the central narrow strip $S\equiv \omega_2\in[\omega_1-\Delta
\omega,\omega_1+\Delta\omega]$), and the second pair
$(\omega_1',\omega_2')$ in another region in the 
$(\omega_1,\omega_2)$ plane that we may want to consider, whereas the
correlation functions $C_{ij}(\br_1,\br_2,\omega_1,\omega_2)$ and
$C_{ij}(\br_1,\br_2,\omega_1',\omega_2')$  themselves
should still be non-zero. Such a situation would signal statistically
independent non-vanishing correlation functions. However, we saw 
that only within a central narrow strip $S$ (whose width is given by
$\kappa_{\rm max}(\eta)$)  in the
$(\omega_1,\omega_2)-$plane 
$C_{ij}$ is non-zero, and within that strip all pairs of
frequencies are used for obtaining a single profile $T(x,y)$.  Here we
show the same thing once more by proving that for $M$ and $V$ to vanish
the second pair of frequencies $\omega_1',\omega_2'$ must be {\em not}
in $S$ --- where, 
however, $C_{ij}(\br_1,\br_2,\omega_1',\omega_2')$ vanishes.

To see this, one first shows with the help of (\ref{eq:umom}) and in a
few lines of calculation that
\begin{eqnarray}
  \label{eq:V2}
  V&=&C_{zz}(\br_1,\br_1,\omega_1,\omega_1')C_{zz}^*(\br_2,\br_2,\omega_2,\omega_2')\,.\\ 
  M&=&C_{zz}(\br_1,\br_2,\omega_1,\omega_2')C_{zz}(\br_1,\br_2,\omega_1',\omega_2)\,.
\end{eqnarray}
We have $C_{zz}(\br_1,\br_1,\omega_1,\omega_1')$ from
\eqref{eq:C3rr2}, where now $\kappa=(\omega_1'-\omega_1)\Delta r/v_s$,
and correspondingly for $C_{zz}(\br_2,\br_2,\omega_2,\omega_2')$.
Whether $V,M$ are large or small can be judged by
comparing it to the product of the standard deviations of each
factor. This corresponds to calculating Pearson's product-moment
coefficients \cite{pearson_note_1895} $V^{\rm res}\equiv
\frac{V}{\sigma(\hat{C})\sigma(\hat{C'})}$ 
and $M^{\rm res}\equiv \frac{M}{\sigma(\hat{C})\sigma(\hat{C'})}$
where we define, for complex $\hat{C}$,
$\sigma(\hat{C})\equiv\sqrt{\sigma^2(\Re\hat{C})+\sigma^2(\Im\hat{C})}$,
and $\hat{C}\equiv \hat{C}_{zz}(\br_1,\br_2,\omega_1,\omega_2)$,
$\hat{C'}\equiv \hat{C}_{zz}(\br_1,\br_2,\omega_1',\omega_2')$  for short.
Going through the same calculation as for $V$, we find after some algebra
\begin{eqnarray}
  \label{eq:sigC2}
\sigma^2(\hat{C})&=& 
C_{zz}(\br_1,\br_1,\omega_1,\omega_1)C_{zz}(\br_2,\br_2,\omega_2,\omega_2)\\
&=& (\pi K_6)^2I^2(0)\,, 
\end{eqnarray}
where 
\begin{equation}
  \label{eq:Iom}
  I(\kappa)\equiv
  \int_{-\infty}^\infty\frac{\tilde{T}_{0,\eta}(\kappa)e^{-|\kappa|\sqrt{\eta^2+\tih^2}}}{\sqrt{\eta^2+\tih^2}}d\eta\,,   
\end{equation}
and hence
$I(0)=\int_{-\infty}^\infty\frac{\tilde{T}_{0,\eta}(0)}{\sqrt{\eta^2+\tih^2}}d\eta$. 
This implies 
\begin{eqnarray}
  \label{eq:Vres}
  |V^{\rm res}|= \left|\frac{I((\omega_1'-\omega_1)\Delta
  r/v_s)I((\omega_2'-\omega_2)\Delta r/v_s)}{I^2(0)}\right|
\end{eqnarray}
For $M$ we have
\begin{eqnarray}
|M^\text{res}|&=&\left| \frac{C_{zz}(\br_1,\br_2,\omega_1,\omega_2')C_{zz}(\br_1,\br_2,\omega_1',\omega_2)}{C_{zz}(\br_1,\br_1,\omega_1,\omega_1)C_{zz}(\br_2,\br_2,\omega_2,\omega_2)} \right|\\
&=&\left|\frac{J(\kappa_{12'},\omega_{c12'})J(\kappa_{1'2},\omega_{c1'2})}{I^2(0)}\right|\, 
\end{eqnarray} 
where
\begin{eqnarray}
  \label{eq:Jofk}
  J(\kappa_{12'},\omega_{c12'})&\equiv& 
  \int\frac{d\eta}{\sqrt{\eta^2+\tih^2}}K(\kappa_{12'}\sqrt{\eta^2+\tih^2},\frac{\Delta 
  r
  \omega_{c12'}}{c}\frac{\eta}{\sqrt{\eta^2+\tih^2}})\tilde{T}_{0,\eta}(\kappa_{12'}) 
\end{eqnarray}
with $\kappa_{12'}\equiv (\omega_2'-\omega_1)/v_s$,
$\kappa_{1'2}\equiv (\omega_2-\omega_1')/v_2$, $\omega_{c12'}\equiv
(\omega_1+\omega_2')/2$, and $\omega_{c1'2}\equiv
(\omega_1'+\omega_2)/2$. From the properties of the integration kernel
$K$ we know that $C_{zz}(\br_1,\br_2,\omega_1,\omega_2')$ vanishes iff
$|\omega_1-\omega_2'|\gtrsim (\omega_1+\omega_2')v_s/(2c\tih)$, and
correspondingly for $C_{zz}(\br_1,\br_2,\omega_1',\omega_2)$. Hence,
for $|M|\ll 1$ and $\omega_{1,2}$ and $\omega_{1,2}'$ all of order
$\omega_0$, we need $|\omega_1-\omega_2'|\gtrsim \omega_0v_s/(c\tih)$
or  $|\omega_1'-\omega_2|\gtrsim \omega_0v_s/(c\tih)$. Note that
$\omega_0v_s/(c\tih)=(1/\chi)v_s/h\gg v_s/h$.   \\

For determining the properties of $V$, we consider our two previous
cases of sources.\\ 
{\em Case 1: Single point source.} Here we have $\tilde{T}$ independent
 of $\kappa$, see eq.\eqref{IPS2}, which inserted into \eqref{eq:Iom} yields
 \begin{equation}
   \label{eq:IokapSPS}
   I(\kappa)=\frac{T_0\Delta
     r}{\sqrt{2\pi}\sqrt{\eta_s^2+\tih^2}}e^{-|\kappa|\sqrt{\eta_s^2+\tih^2}}
 \end{equation}
and hence
\begin{equation}
  \label{eq:VresSPS}
  |V^\text{res}|=e^{-\frac{\Delta
      r}{v_s}\sqrt{\eta_s^2+\tih^2}(|\omega_1'-\omega_1|+|\omega_2'-\omega_2|)}\,. 
\end{equation}
For sources at $\eta_s\sim \tih$, we have therefore
$|V^\text{res}|\ll 1$ iff $|\omega_1'-\omega_1|>\delta \omega$ or
$|\omega_2'-\omega_2|>\delta \omega$ where $\delta\omega\equiv
v_s/(\Delta r \sqrt{\eta_s^2+\tih^2})\sim v_s/(\Delta r
\tih)=v_s/h\sim 10^{-2}$\,Hz.  

{\em Case 2: Constant temperature field in the positive upper half
  plane}. Here, $\tilde{T}(\kappa_x,\eta)$ is
given by eq.\eqref{eq:Ttiluni}. Hence, $V^{\rm res}=0$ as soon as
$\omega_1'\ne \omega_1$ or  $\omega_2'\ne \omega_2$. Of course, the
$\delta(\kappa_x)$ function in \eqref{eq:Ttiluni} 
arises from the complete lack of structure of the temperature profile
in $x$-direction. More realistic is at least a cut-off at the size of
Earth, which we take as the same as in $y$ direction. In that case one
finds $\tilde{T}_{\br'}(\kappa_x)=(\hat{y} T_0/\pi)\sinc \left(\frac{\kappa_x
    \hat{y}}{\pi }\right)$ and hence 
\begin{equation}
  \label{eq:IkappSPS}
I(\kappa)=(\hat{y} T_0/\pi)\sinc \left(\frac{\kappa_x
    \hat{y}}{\pi
  }\right)\int_0^{\hat{\eta}}\frac{d\eta}{\sqrt{\eta^2+\tih^2}}e^{-|\kappa|\sqrt{\eta^2+\tih^2}}\,. 
\end{equation}
The exponential factor in the integral makes again that $V^{\rm res}$
 vanishes essentially if 
$|\omega_i'-\omega_i| \gtrsim v_s/h$ for $i=1$ or $i=2$. \\

Comparing with the
situation for $M$, we find that for both types of sources considered,
$V$ vanishes much more rapidly as 
function of the separation of two frequencies, as there is no factor
$1/\chi$ multiplying $v_s/h$. Hence, the request for vanishing $M$ is
more restrictive.\\

The question of the usefulness
of considering other frequency pairs can now be phrased as: Can one find pairs
of frequencies $(\omega_1',\omega_2')$ such that
$|\omega_2'-\omega_1|\gg\Delta \omega\equiv (1/\chi)v_s/h$ or
$|\omega_1'-\omega_2|\gg\Delta 
\omega$ while still $|\omega_2'-\omega_1'|\lesssim\Delta\omega$, {\em for all}
frequencies $\omega_1,\omega_2$ with
$|\omega_2-\omega_1|\lesssim\Delta\omega$ used
in the reconstruction of a temperature profile from
$C(\br_1,\br_2,\omega_1,\omega_2)$? 
For a single frequency pair $(\omega_1',\omega_2')$ all conditions can
be easily satisfied. It is enough that both pairs
$(\omega_1,\omega_2)$ and $(\omega_1',\omega_2')$ be inside the strip
$S$, and at the same time far away from each other, i.e.~$|\omega_1-\omega_1'|\gg \Delta\omega$, which implies 
$|\omega_2'-\omega_1|\gg\Delta \omega$ and
$|\omega_1'-\omega_2|\gg\Delta \omega$ at the same time.  However, the
difficulty arises from the fact that we use already
all pairs $(\omega_1,\omega_2)$ in the full available band-width for
the reconstruction of a single temperature profile.  This can be seen
e.g.~from eq.\eqref{eq:FC}, where we integrate over all
$\tilde{k}_c=\Delta r(\omega_1+\omega_2)/(2c)$ for recovering $\tilde{T}$.
Hence, there are really no new frequency pairs that can be used for
improving the signal/noise ratio of the reconstructed temperature
profile. 

The same conclusion can be arrived at more formally by calculating the
correlations between temperature profiles obtained from different
center frequencies. Let $T_\text{rec}(x,y;\omega_0)$  be the
reconstructed temperature profile given by eq.\eqref{eq:Trec}, where
we now keep explicit the dependence on the center frequency
$\omega_0$, hidden in that equation in the filter functions
$A(\omega_1,\omega_0)$, see eqs.(\ref{eq:CG},\ref{eq:CG2}), and
$C_{ii}^F\to\hat{C}_{ii}^F$ is understood, so as to get the
temperature profile from a single realization of the noise process. We
define the correlation function
\begin{equation}
  \label{eq:KT}
 K(T_\text{rec}(\omega_{01}),T_\text{rec}(\omega_{02}))\equiv\langle T_\text{rec}(x,y;\omega_{01})T_\text{rec}(x,y;\omega_{02})\rangle-\langle T_\text{rec}(x,y;\omega_{01})\rangle\langle T_\text{rec}(x,y;\omega_{02})\rangle\,, 
\end{equation}
and its renormalized dimensionless version 
\begin{equation}
  \label{eq:KTrel}
  K_\text{rel}(T_\text{rec}(\omega_{01}),T_\text{rec}(\omega_{02}))\equiv
  K(T_\text{rec}(\omega_{01}),T_\text{rec}(\omega_{02}))/K(T_\text{rec}(\omega_{01}),T_\text{rec}(\omega_{01}))
\end{equation}
that obviously satisfies
$K_\text{rel}(T_\text{rec}(\omega_{01}),T_\text{rec}(\omega_{01}))=1$. We
have
\begin{align}
  \label{eq:KTa}
  K(T_\text{rec}(\omega_{01}),T_\text{rec}(\omega_{02}))&=
\left(\frac{1}{2\pi K_6\Delta r}\right)^2\int d\kappa_1\,d\kappa_2\,d\tik_{c1}\, d\tik_{c2}\frac{\sqrt{|\tik_{c1}\tik_{c2}|}}{2\pi}\nonumber\\  
&\times e^{-i(\kappa_1-\kappa_2)(\tix-\tix_0)}e^{i(\tik_{c1}-\tik_{c2})\zeta}F(\tik_{c1},\tik_{c0}^{(1)},\frac{\tib}{\sqrt{2}})F(\tik_{c1},\tik_{c0}^{(2)},\frac{\tib}{\sqrt{2}})\nonumber\\
&  \times \big(\langle\hat{C}_{zz}(\br_1,\br_2,\omega_1,\omega_2)\hat{C}_{zz}^*(\br_1,\br_2,\omega_1',\omega_2')\rangle\nonumber\\
&-\langle\hat{C}_{zz}^*(\br_1,\br_2,\omega_1,\omega_2)\rangle\langle\hat{C}_{zz}(\br_1,\br_2,\omega_1',\omega_2')\rangle\big)\,,  
\end{align}
where $\kappa_1=(\omega_2-\omega_1)\Delta r/v_s$, $\kappa_2=(\omega_2'-\omega_1')\Delta r/v_s$, $\tik_{c1}=(\omega_1+\omega_2)\Delta r/(2c)$, $\tik_{c2}=(\omega_1'+\omega_2')\Delta r/(2c)$, $\tik_{c0}^{(i)}=\omega_0^{(i)}$ ($i=1,2$), $F(\tik_c,\tik_{c0},\frac{\tib}{\sqrt{2}})=A(\omega_1,\omega_0)A^*(\omega_2,\omega_0)$ with $\tik_{c0}=\Delta r \omega_0/c$ (see eq.\eqref{eq:A}), and we have used that $T_\text{rec}\in\mathbb R$.       
We evaluate  $K(T_\text{rec}(\omega_{01}),T_\text{rec}(\omega_{02}))$
for the case of constant temperature in the upper half plane.  Using
\eqref{eq:V}, \eqref{eq:Cuni}, and switching momentarily to
integration variables $\omega_1,\omega_2,\omega_1',\omega_2'$, and
then back to $\kappa_1\,\tik_{c1}$, we are led to 
\begin{eqnarray}
  \label{eq:KTb}
  K(T_\text{rec}(\omega_{01}),T_\text{rec}(\omega_{02}))&=&
\frac{T_0^2v_s}{2\pi c}\int d\kappa_1\,d\tik_{c1}|\tik_{c1}|
e^{2i\zeta\tik_{c1}}F(\tik_{c1},\tik_{c0}^{(1)},\frac{\tib}{\sqrt{2}})F(\tik_{c1},\tik_{c0}^{(2)},\frac{\tib}{\sqrt{2}})\,.    
\end{eqnarray}
The integral is clearly real, as it should. The integral over
$\kappa_1$ leads, when integrated from $-\infty$ to $\infty$ to a
divergent factor, but that factor cancels (together with the remaining
prefactor $T_0^2v_s/(2\pi c)$) when we consider the re-scaled version
of the correlation function $
K_\text{rel}(T_\text{rec}(\omega_{01}),T_\text{rec}(\omega_{02}))$. If
we set $\tik_{c0}^{(2)}=\tik_{c0}^{(1)}+\delta\tik_{c0}$, it is clear
that the only remaining scale for $\delta\tik_{c0}$ is
$\tib/\sqrt{2}$. The remaining integral over $\tik_{c1}$ in \eqref{eq:KTb}
can in fact be evaluated analytically.  The result is too cumbersome
to be reported here, but plotting it as function of $\delta\tik_{c0}$
shows that indeed the correlations decay only on a scale of order
$\tib$. This proves that by shifting the center frequency within the
available bandwidth one cannot gain independent estimates of
$T_\text{rec}$ that would allow one to improve substantially the
signal-to-noise ratio.

\subsection{Additional antennas}
So far we considered only two antennas.  As mentioned before, in order
to obtain a reconstructed 
single source image with a single peak, one may sum the correlated
signals from several antenna pairs.  It is to be expected that this
will reduce $\sigma(T_{\rm rec})/T_{\rm  rec}$, but we have to figure
out how far two pairs of antenna have to be separated in order to
produce essentially uncorrelated correlation functions.  To answer
this question, we
have to generalize eq.\eqref{eq:V} to pairs of
correlators at different points $\br_1',\br_2'$.
We define
\begin{equation}
  \label{eq:Vr}
  V_r\equiv \langle
  \hat{C}(\br_1,\br_2,\omega_1,\omega_2)\hat{C}^*(\br_3,\br_4,\omega_1,\omega_2)\rangle - \langle
  \hat{C}(\br_1,\br_2,\omega_1,\omega_2)\rangle\langle\hat{C}^*(\br_3,\br_4,\omega_1,\omega_2)\rangle\,, 
\end{equation}
where we take $\br_{i+2}=\br_i+\rho_i\hat{e}_y$ for $i=1,2$, i.e.~the
antennas in the second pair are shifted by  distance $\rho_i$ in
$y$-direction 
compared to the corresponding ones in the first pair. From
\eqref{eq:umom} we have 
\begin{equation}
  \label{eq:Vr2}
  V_r=C(\br_1,\br_3,\omega_1,\omega_1)C(\br_4,\br_2,\omega_2,\omega_2)\,,
\end{equation}
where from \eqref{Cfin} and \eqref{eq:K0}
\begin{eqnarray}
  \label{eq:Cww}
  C(\br_1,\br_3,\omega_1,\omega_1)&=&K_6\int\frac{d\eta}{\sqrt{\eta^2+\tih^2}}K(0,\frac{
  \rho_1\omega_1}{c}\frac{\eta}{\sqrt{\eta^2+\tih^2}})\tilde{T}(0,\eta)\\
&=&K_6\int\frac{d\zeta}{1-\zeta^2}(J_0(\frac{\rho_1\omega_1\zeta}{c})-iH_0(\frac{\rho_1\omega_1\zeta}{c}))\,.
\end{eqnarray}
The corresponding result for $C(\br_4,\br_2,\omega_2,\omega_2)$ is
obtained from the last line in \eqref{eq:Cww} by replacing
$\rho_1\to\rho_2$. 
For the uniform temperature field in the upper half plane up to a
cut-off $\hat{y}$ and also a cut-off of the same value in
$x$-direction, we have $T(0,\eta)=\hat{y}T_0/\pi$ for $0\le y\le
\hat{y}$. No closed form was found for the remaining integral over
$\zeta$, but a closed form is easily obtained if we neglect the slowly
varying envelope $1/(1-\zeta^2)$, which is legitimate for cut-offs
$\hat{\eta}$ not too close to 1 and gives an idea on which
length-scale $V_r$ will vanish.  Plotting the results of the
integration one finds that both real and imaginary part decay on a
scale of $\rho_i\omega_0/c\sim 1$, where we have used again
$\omega_1\simeq \omega_2\simeq \omega_0$.  Hence, for the correlation
functions of two pairs of antenna to decorrelate, it is enough that
one antenna in one pair be at a distance of order $r\gtrsim
c/\omega_0=\lambda/2\pi$,  
i.e.~of the order of the central wave-length 
$\lambda$ with respect to at least one antenna of the other pair.
For standard parameters, this is of the order 10\,cm, 
neglecting factors of order 1 (the $2\pi$ helps of course, but for the
imaginary part of $C(\br_1,\br_2,\omega_1,\omega_1)$ there is a
comparable factor in the scale). Extending the separation of the two
antennas in the original pair to $2\Delta r=200$\,m, one would have
place for about 
2000 antennas in between. This in turn would then allow to be built
correlations from $10^6$ pairs of antenna, where the antennas
in each pair are still separated at least by $\Delta
r=100$\,m. Considering that averaging of $N$ temperature profiles
obtained from $N$ statistically independent correlation functions
improves the signal-to-noise ratio of the average temperature profile
by a factor $\sqrt{N}$, we can improve the SNR by a factor $10^3$. If
considering the prefactors of order one, 10 times more antennas can
be used, the SNR could be improved by a factor $10^4$.  But even
such a large improvement is not yet sufficient to beat the low SNR
of order $\chi^{3/2}\simeq 10^{-5}$.  It is quite likely, however,
that a displacement of an antenna also in $x$ direction by a distance
of the order $\lambda/(2\pi)$ leads to a completely decorrelated
correlation function.  If so, one might gain another factor  up to $10^3$
in the SNR by considering quasi-1D antenna arrangements, with a 
width in $x$-direction of order 10 meters.  In the latter case one
should then be able (after averaging temperature profiles obtained
from some $10^{14}$ correlation functions from that many pairs of
antennas) to achieve an SNR of $10^2$, and hence a RR of order of a
few Kelvin.  However, it is obvious that the effort for doing so is
humungous, and the same geometrical and radiometric resolution might
be achievable more easily with other means.  \\

Other interesting
ideas of improving the SNR involve using focussing antennas for
increasing the flux, and/or exploiting higher order correlation
functions as well, but these are beyond the scope of the present
investigation.  \\

\section{Discussion}
We examined the fundamental feasibility  of a
new type of  passive remote microwave-imaging of a 2D scenery with a
satellite 
having only a 1D antenna array, 
arranged %
perpendicular to the direction of flight of the
satellite.  We analyzed the simplest possible configuration of only
two antennas. 
The scheme  
is based on correlating Fourier components of the observed electric field
fluctuations at the position of the two antennas at slightly different
frequencies $\omega_1$ and 
$\omega_2$, and leads 
effectively to a mapping of the 2D brightness temperature as function of
position $x,y$ to  correlations as function of the center
frequency 
$\omega_c=(\omega_1+\omega_2)/2$ and the frequency difference 
$\Delta\omega=\omega_1-\omega_2$.  With
two antennas separated by $\Delta r$, center frequency $\omega_c$ and a
satellite flying at height $h$, the resolution both in $x$-
and $y$-direction is of the order
$h\chi=hc/(\Delta r \omega_c)$.  Only very small frequency-differences %
lead to correlations of finite, useful magnitude. For typical intended
SMOS-NEXT values they 
are of the order of at most 10Hz, which, however, still have to be
divided by the number of points in $x$ direction that one wants to
resolve within a snapshot. 
This implies that one must be able to measure GHz frequencies with accuracy of
the order of a 1/10-1/100 Hz.
The speed $v_s$ 
of the satellite only enters in the maximum frequency difference useful for
correlating the signals, which is given by $\Delta \omega\lesssim (\Delta
r/h)(v_s/c)\omega_c$. 

In the minimal situation of two antennas,  the relative radiometric
resolution 
$\sigma(T)/T$  is, for a single point source of order
$\sqrt{\chi}$, whereas 
for a uniform temperature field in the positive half plane $y>0$ (up
to some large cut-off of the size of Earth),
$\sigma(T_\text{rec})/T_\text{rec}\sim 
1/\chi^{3/2}$ which for standard parameters is of order $10^5$.  
We have neglected so far the additional noise that comes from the
antennas themselves, such that our results should be considered as
lower bounds for $\sigma(T_\text{rec})/T_\text{rec}$. 
Unfortunately, this large uncertainty prevents a direct application of the
method with just two antennas, and massive noise reduction is
required. Some ideas are 
discussed in Sec.\ref{sec.noisered}, where it was found that one can
obtain statistically independent correlation functions by displacing
one antenna by a distance of the order of $\lambda/2\pi$, where
$\lambda$ is the central wave-length. Hence, the signal-to-noise ratio
$T_\text{rec}/\sigma(T_\text{rec})$ can be massively increased by a
factor $N$ when using the correlations from $\sim N(N-1)/2$ pairs of 
antennas from $N$ antennas separated all by at least a distance of
order $\lambda/2\pi$. However, the computational effort and the size
of the overall structure appear forbiddingly large for
achieving a radiometric resolution of order of a few Kelvin with a
geometrical resolution of order one kilometer.\\

An alternative application might
be the precise localization of very strong point sources that by far
dominate the more or less uniform background from Earth's thermal
emission. As long as one is not interested in a very precise
measurement of the intensity of the source, one might localize it very
precisely using just two widely separated antennas. These need not
even by an board of the same satellite.  By having two satellites with
well-known distance separated by about 100km for instance, the
geometrical resolution 
achievable in the microwave regime would be of the order of a meter in
both $x-$ and $y-$ direction and with rather small computational effort,
opening interesting perspectives for such applications. It should also
be kept in mind that the method can be easily transferred to other
types of waves, sources, and media.  For example 2D ultra-sound
imaging might be possible by beating the signals of just two moving
microphons.  Different physical systems can be easily mapped to each  
other by comparing the corresponding dimensionless parameters
introduced in Sec.\ref{sec.intro}.

\section{Acknowledgments}
The authors are very thankful to the
CESBIO SMOS team and the CNES project management, without whom
this work would not have succeeded. We have more particularly
appreciated the technical support from Fran\c{c}ois Cabot, Eric 
Anterrieu, Ali Khazaal, Guy Lesthi{\'e}vant, and Linda Tomasini.

\section{Appendix}
\subsection{Current fluctuations and temperature}
The connection between the intensity of the current fluctuations and
the local temperature can be found e.g.~in
\cite{sharkov_passive_2003,rytov_theory_1959,Rytov89,carminati_near-field_1999}.   
For being self-contained and relating to the notations used in this
paper, we here give a short derivation of this connection. We also
show that $\hat{E}_{z,\br}(\omega)$ is, in the frequency range
considered, a circularly symmetric Gaussian process.

\subsubsection{Thermal radiation}\label{sec:thrad}
We begin by recalling the energy density of electromagnetic black body
radiation at frequency $\omega$, $u(\omega)=\hbar\omega\rho(\omega)f(\omega,T)$,
where 
$\rho(\omega)=\omega^2/(\pi^2c^3)$ is the density of states (number of modes
between frequencies $\omega$ and $\omega+d\omega$ per volume), and
$f(\omega,T)=1/(e^{\hbar\omega/(k_BT)}-1)$ the thermal Bose occupation
factor, with $k_B$ the Boltzmann constant, and $T$ the absolute
temperature of the 
radiation field. An infinitesimal patch on the surface at position
$x,y$ with surface
$dA$ and temperature $T(x,y)$ in thermal equilibrium with the radiation
field in its immediate vicinity,
radiates off an amount of energy per unit time and at frequency $\omega$
given by $dA\, u\,c \cos\theta$ in direction $\theta$ with respect to
the 
the surface normal.  The
energy density for 
both polarization directions received at the position of the satellite
at distance $R$ from this patch also varies $\propto \cos\theta$, and
energy 
conservation requires
\begin{equation}
  \label{eq:dus}
  du_s(\omega)=\frac{dA u(\omega)\cos\theta}{2\pi R^2}=\frac{dA \hbar
    \omega^3\cos\theta}{2\pi^3c^3R^2(e^{\hbar\omega/(k_BT)}-1 )}\,.
\end{equation}
Earth is rather a grey than a black body, and we therefore have to
include the emissivitiy of the patch $B(x,y;\omega,\theta,\varphi)$ 
in the direction of the satellite given by polar and azimuthal
angles. It can also depend on polarization, which we skip here for
simplicity. Integration over the whole radiating surface 
gives the entire energy 
 density at the position of the satellite at this frequency, 
\begin{equation}
  \label{eq:us}
  u_s(\omega)=\int \frac{dx\,dy \, u}{2\pi R^2}=\int\frac{dx\,dy \hbar
    \omega^3\cos\theta(x,y,h)B(x,y;\omega,\theta,\phi)}{2\pi^3c^3(h^2+x^2+y^2)(e^{\hbar\omega/(k_BT(x,y))}-1 
    )}\,.  
\end{equation}
In the microwave regime and temperatures $T\simeq 300K$, $\hbar\omega$ is
about four orders of magnitude smaller than $k_BT$, such that the Bose
factor becomes, to first order in $\hbar\omega/k_B T$, $f(\omega,T)\simeq
k_B T/(\hbar\omega)$, with corrections of order $10^{-4}$. This simplifies
$u_s$ to 
\begin{equation}
  \label{eq:us1}
  u_s(\omega)=\frac{k_B}{2\pi^3c^3}\int  dx\,dy\, \omega^2 \frac{T_B(x,y)\cos\theta(x,y,h)}{h^2+x^2+y^2}\,,
\end{equation}
where we defined the brightness temperature $T_B(x,y)\equiv
T(x,y)B(x,y;\omega,\hat{k})$, i.e.~the
absolute temperature 
a black body would need to have for producing the same thermal
radiation intensity at the frequency and in the direction $\hat{k}$
considered, specified explicitly by the two angles $(\theta,\varphi)$. 
At the same time, the total energy density (integrated over all frequencies)
at  position $\br_1$ of antenna 1 is $U_s=\int d\omega
u_s(\omega)=\frac{\epsilon_0}{2}\langle 
\bE^2(\br_1)\rangle$, where 
the average is over the thermal ensemble, but due to ergodicity
we may also average over time,
$\langle\ldots\rangle_{\tau_a}=\frac{1}{\tau_a}
\int_{-\tau_a/2}^{\tau_a/2}(\ldots) dt$. In the end one should take
the limit $\tau_a\to\infty$. In fact, we may even average over both the
thermal ensemble and time,
i.e.~$U_s=\frac{\epsilon_0}{2}\langle\langle\bE^2(\br_1)\rangle\rangle_{\tau_a}$ 
Expressing
then the electric field in 
terms of its Fourier transform, the time integral leads to a sinc-function,  
\begin{equation}
  \label{eq:Us}
  U_s=\frac{\epsilon_0}{4\pi}\int\int
  \sinc\big((\omega'-\omega)\frac{\tau_a}{2\pi}\big)\langle
  \tilde{\bE}_{\br_1}^*(\omega)\tilde{\bE}_{\br_1}(\omega')\rangle
  d\omega\,d\omega'\,,  
\end{equation}
  with $\sinc(x)\equiv \sin(\pi x)/(\pi x)$. For large $\tau_a$, the
  sinc-function can be replaced by
  $(2\pi/\tau_a)\delta(\omega-\omega')$, and we are then left with 
\begin{equation}
  \label{eq:Us1}
  U_s=\frac{\epsilon_0}{2\tau_a}\int \langle
 | \tilde{\bE}_{\br_1}(\omega)|^2\rangle
  d\omega\,.
\end{equation}
Therefore, the energy density per unit frequency  at frequency omega
is given by $u_s(\omega)=\frac{\epsilon_0}{2\tau_a}\langle
 | \tilde{\bE}_{\br_1}(\omega)|^2\rangle$. Together with eq.(\ref{eq:us1})
 we thus have 
\begin{equation}
  \label{eq:E2}
\langle
 |\tilde{\bE}_{\br_1}(\omega)|^2\rangle=\frac{\tau_ak_B}{\pi^3\epsilon_0c^3}
 \int dx\,dy\,\frac{\omega^2T_B(x,y)\cos\theta(x,y,h)}{h^2+x^2+y^2}\,.
\end{equation}
The connection to the current fluctuations is found by comparing this
expression to what we obtain from eq.(\ref{C}) if we do not  use
(\ref{eq:jtoT}) yet.  There we set $i=j$, 
$\br_1=\br_2=(0,0,h)$, $\omega_1=\omega_2$, and $v_s=0$, as we are interested in the
energy density in a given fixed point $\br_1$, identical to the original
position of the antenna.  This gives
\begin{eqnarray}
  \label{eq:E12}
\langle
 |\tilde{E}_{i,\br_1}(\omega)|^2\rangle  
&=&K_4\int_{-\infty}^\infty
dt_1\int_{-\infty}^\infty
dt_2 \int_{-\infty}^\infty d\omega'\int
dx\,dy\frac{\omega'^2\langle|\tilde{j_i}(x,y,\omega')|^2\rangle}{h^2+x^2+y^2}\nonumber\\   
&&\times
e^{i(\omega-\omega')(t_1-t_2)}\,, 
\end{eqnarray}
where $K_4=K_1^2l_c^3d/(4\pi^2\tau_c)$, and we have already restricted
the current density to the surface of Earth, i.e.~assumed
$\langle|\tilde{j_i}(\br',\omega')|^2\rangle=d \langle|\tilde{j_i}(x,y,\omega')|^2\rangle\delta(z)$. In practice, the time
integrals originating from the Fourier transforms will be taken over a
finite time $\tau_F$. Since the only time-dependence is in the exponential,
the time-integrals can be done
exactly, leading to 
\begin{equation}
  \label{eq:I1}
  \int_{-\tau_F/2}^{\tau_F/2}\int_{-\tau_F/2}^{\tau_F/2}dt_1\,dt_2
  e^{i(\omega-\omega')(t_1-t_2)}=
  \frac{2}{(\omega-\omega')^2}\big(1-\cos(\tau_F(\omega-\omega'))\big)\,.
\end{equation}
For large $\tau_F$, this function is highly peaked at $\omega=\omega'$ and
behaves as $2\pi \tau_F\delta(\omega-\omega')$, where the prefactor may be
verified by integrating over. We are thus led to
\begin{equation}
  \label{eq:E22}
\langle
 |\tilde{E}_{i,\br_1}(\omega)|^2\rangle  
=K_42\pi\tau_F\int
dx\,dy\frac{\omega^2\langle|\tilde{j_i}(x,y,\omega)|^2\rangle}{h^2+x^2+y^2}\,.
\end{equation}
The thermal fluctuations of the electric field are isotropic in their
intensity, such that one third of the energy is in a given polarization
direction $i$, i.e.~$\langle
 |\tilde{E}_{i,\br_1}(\omega)|^2\rangle=\frac{1}{3}\langle
 |\tilde{\bE}_{\br_1}(\omega)|^2\rangle$.  Inserting eq.(\ref{eq:E2}) for
 the latter quantity, we are led to 
\begin{equation}
  \label{eq:E23}
 \langle
 |\tilde{E}_{i,\br_1}(\omega)|^2\rangle  
=\frac{\tau_ak_B}{3\pi^3\epsilon_0c^3}
 \int dx\,dy\,\frac{\omega^2T_B(x,y)\cos\theta(x,y,h)}{h^2+x^2+y^2}\,.
\end{equation}
Comparison with eq.(\ref{eq:E22}) allows one to identify
\begin{equation}
  \label{eq:jtoT2}
\langle|\tilde{j_i}(x,y,\omega)|^2\rangle=K_2 T_\text{eff}(x,y), 
\end{equation}
with $K_2=32\tau_a\tau_ck_B/(3 \tau_Fl_c^3d \mu_0 c)$ and
$T_\text{eff}(x,y)\equiv T_B(x,y)\cos\theta(x,y,h)$.
Thus, the current fluctuations are given directly by the brightness
temperature (rescaled by the directional-$\cos\theta(x,y,h)$), up to a
constant prefactor.  As mentioned in the Introduction, we write  
$T$ for short for $T_\text{eff}$ in the rest of the article.  The
constant prefactor depends 
on the time intervals for 
averaging and the Fourier transforms, but in the end we will always be 
interested in relative radiometric resolution,
i.e.~$\sigma(T(x,y))/T(x,y)$, 
where $\sigma(T(x,y))$ denotes the standard deviation of the reconstructed
temperatures over the thermal ensemble of the radiation field, such that
the constant prefactor cancels out.   \\


\subsubsection{Circular symmetry}\label{sec.circular}
A Gaussian distribution of a complex jointly-Gaussian random vector
$z=(z_1,z_2,\ldots,z_n)\in\mathbb{C}^n$ is fully characterized 
by the  expectation
values, $E(z_i)\,\forall i$, the covariance matrix $K=E[z\,z^\dagger]$, and
the pseudo-covariance 
matrix $M=E[z\,z^t]$. Both matrices together specify the correlations
between the four different combinations of real and imaginary parts of
the $z_i$. The Gaussian distribution is called circularly symmetric,
if $P(z)$ is invariant under the transformation $z\mapsto z e^{i\phi}$
with an arbitrary real phase $\phi$. One shows that a distribution is
Gaussian symmetric if and only if $M=0$. This implies immediately that
$E[z_i]=0\,\forall i$ \cite{Gallager08}. \\

The corresponding definitions and statements for complex Gaussian
processes are easily obtained by replacing the discrete index $i$ in
$z_i$ by a continuous one, e.g.~a time argument, or in our case of
$\hat{E}_{z,\br}(\omega)$, a 4-component real vector with
a ``continuous index'' $\omega,\br$. In order to show that
$\hat{E}_{z,\br}(\omega)$ is a circularly symmetric complex Gaussian
process over $\omega,\br$, we need to prove that 
$0=M(\br_1,\br_2,\omega_1,\omega_2)\equiv\langle\hat{E}_{z,\br_1}(\omega_1)\hat{E}_{z,\br_2}(\omega_2)\rangle$,
at least in the narrow frequency band that we are interested in. In
view of eq.(\ref{jtoE}), for 
this it is enough to show that
$M_J\equiv\langle\tilde{j}_{z,\br_1}(\omega_1)\tilde{j}_{z,\br_2}(\omega_2)\rangle
=0$.  Expressed as Fourier transforms of the time-dependent
current-densities, this correlator equals
\begin{equation}
  \label{eq:jtjt}
 \langle\tilde{j}_{z,\br_1}(\omega_1)\tilde{j}_{z,\br_2}(\omega_2)\rangle
=\frac{1}{2\pi}\int_{-\infty}^{\infty}dt_1\,dt_2 e^{-i(\omega_1
  t_1+\omega_2 t_2)}\langle
j_{z,\br_1}(t_1)j_{z,\br_2}(t_2)\rangle\,.
\end{equation}
The physical origin of the current fluctuations are thermal
fluctuations, and the condition of thermal equilibrium implies that
the current correlator is invariant under global time-translation
(i.e.~a shift of the origin of the time axis of $t_1$ and $t_2$ by the
same amount) and hence depends only on $t_2-t_1$, $\langle
j_{z,\br_1}(t_1)j_{z,\br_2}(t_2)\rangle=f(\br_1,\br_2,\tau)$,
where $\tau=t_2-t_1$, and we will also use $t=(t_2+t_1)/2$. With this we get
  \begin{eqnarray}
    \label{eq:MJ}
    M_J&=&\frac{1}{2\pi}\int_{-\infty}^\infty
    dt\,e^{-i(\omega_1+\omega_2)t}\int_{-\infty}^\infty
    d\tau\,e^{-i(\omega_2-\omega_1)\tau/2}f(\br_1,\br_2,\tau)\nonumber\\
&=&\sqrt{2\pi}\delta(\omega_1+\omega_2)\tilde{f}(\br_1,\br_2,(\omega_2-\omega_1)/2)\,,
  \end{eqnarray}
where $\tilde{f}(\br_1,\br_2,\omega)$ is the Fourier transform of
$f(\br_1,\br_2,t)$ with respect to time. The $\delta$-function
implies that $M_J$ vanishes unless $\omega_1=-\omega_2$. But we are
interested only in frequencies in the small interval $\omega_2\in
[\omega_1-\Delta \omega,\omega_1+\Delta \omega]$, centered close
to $\omega_0$ of the order of 1.4GHz. Hence, in this frequency
interval we have indeed $M_J=0$, and the complex Gaussian process 
$\hat{E}_{z,\br}(\omega)$ over $\omega,\br$ can be considered as
circularly symmetric and eq.\eqref{eq:1324} valid. In particular, it
does not contain any correlator of the type $E\,E$, but only of the
type $E\, E^*$.


\bibliography{../../../../../mypapers/mybibs_bt_280417bkp}

\begin{thebibliography}{10}
\providecommand{\url}[1]{\texttt{#1}}
\providecommand{\urlprefix}{URL }
\providecommand{\eprint}[2][]{\url{#2}}

\bibitem{Interferometry_Synthesis_Radio_Astronomy}
Thompson, A.~R., Moran, J.~M., \& George W.~Swenson, J., \emph{Interferometry
  and Synthesis in Radio Astronomy, 2nd Edition} (WILEY-VCH, 2001).

\bibitem{Kerr2001}
Kerr, Y.~H. \emph{et~al.}, Soil Moisture Retrieval from Space: The Soil
  Moisture and Ocean Salinity (SMOS) Mission, \emph{IEEE Transactions on
  geoscience and remote sensing} \textbf{39 (8)} (2001).

\bibitem{Kerr2001b}
Kerr, Y. \emph{et~al.}, Overview of SMOS performance in terms of global soil
  moisture monitoring after six years in operation, \emph{Remote Sensing of
  Environment} \textbf{180}, 40--63 (2016).

\bibitem{Kerr2010}
Kerr, Y.~H. \emph{et~al.}, The SMOS Mission: New Tool for Monitoring Key
  Elements of the Global Water Cycle, \emph{Proceedings of the Ieee}
  \textbf{98(5)}, 666--687 (2010).

\bibitem{camps_two-dimensional_2001}
Camps, A. \& Swift, C., A two-dimensional {Doppler}-{Radiometer} for {Earth}
  observation, \emph{IEEE Transactions on Geoscience and Remote Sensing}
  \textbf{39}, 1566--1572 (2001).

\bibitem{braun_generalization_2016}
Braun, D., Monjid, Y., Roug\'e, B., \& Kerr, Y., Generalization of the {Van}
  {Cittert}–{Zernike} theorem: observers moving with respect to sources,
  \emph{Meas. Sci. Technol.} \textbf{27}, 015002 (2016).

\bibitem{Jackson99}
Jackson, J., \emph{Classical Electrodynamics}, 3rd.edition (Wiley, 1999).

\bibitem{Kerr2012}
Kerr, Y.~H. \emph{et~al.}, The SMOS Soil Moisture Retrieval Algorithm,
  \emph{Ieee Transactions on Geoscience and Remote Sensing} \textbf{50(5)},
  1384--1403 (2012).

\bibitem{sharkov_passive_2003}
Sharkov, E.~A., \emph{Passive {Microwave} {Remote} {Sensing} of the {Earth}:
  {Physical} {Foundations}} (Springer, Berlin ; New York : Chichester, UK,
  2003), 2003 edition edn.

\bibitem{LLStatMech80}
Landau, L.~D., Lifshitz, E.~M., \& Pitaevskii, L.~P., \emph{Statistical
  Physics} (Pergamon Press, Oxford, 1980), 13 edn.

\bibitem{rytov_theory_1959}
Rytov, S.~M. \& ~, Herman, E., \emph{Theory of electric fluctuations and
  thermal radiation} (Electronics Research Directorate, Air Force Cambridge
  Research Center, Air Research and Development Command, U.S. Air Force],
  Bedford, Mass., 1959).

\bibitem{Rytov89}
Rytov, S.~M., Kravtsov, Y.~A., \& Tatarskii, V.~I., \emph{Principles of
  Statistical Radiophysics} (Springer-Verlag, Berlin, 1989).

\bibitem{carminati_near-field_1999}
Carminati, R. \& Greffet, J.-J., Near-Field Effects in Spatial Coherence of
  Thermal Sources, \emph{Phys. Rev. Lett.} \textbf{82}, 1660--1663 (1999).

\bibitem{NumPaper}
Monjid, Y., Roug\'e, B., Kerr, Y., \& Braun, D., to be published .

\bibitem{Barrett03}
Barrett, H.~H. \& Myers, K.~J., \emph{Foundations of Image Science} (Wiley,
  2003).

\bibitem{pearson_note_1895}
Pearson, K., Note on {Regression} and {Inheritance} in the {Case} of {Two}
  {Parents}, \emph{Proc. R. Soc. Lond.} \textbf{58}, 240--242 (1895).

\bibitem{Gallager08}
Gallager, R.~G., Circularly-Symmetric Gaussian random vectors (2008),
  \eprint{unpublished}.

\end{thebibliography}
\end{document}